\documentclass[12pt,a4paper,hyper]{JHEP3}
\usepackage{epsfig}
\usepackage{amsmath,amsfonts,amssymb}
\usepackage{multirow}
\usepackage{slashed}
\usepackage{graphicx}
\usepackage{epsfig}
\usepackage{psfrag}
\usepackage{color}

\usepackage{ulem}

\def\={\; = \;}

\newenvironment{myitemize}{
\begin{itemize}
\setlength{\itemsep}{1pt}
\setlength{\parskip}{0pt}
\setlength{\parsep}{0pt}}{\end{itemize}}
\newcommand{\bem}{\begin{pmatrix}}
\newcommand{\eem}{\end{pmatrix}}

\def\h{\eta}

\def\O{{\Omega}}

\def\be{\begin{equation}}
\def\ee{\end{equation}}
\def\ba{\begin{align}}
\def\ea{\end{align}}
\def\bse{\begin{subequations}}
\def\ese{\end{subequations}}
\def\1F1{{}_1\!F_1}
\def\2F0{{}_2\!F_0}

\def\h3{$\textrm{H}_3^+$}
\font\manual=manfnt
\def\dbend{\lower3.5pt\hbox{\manual\char127}}

\def\bar{\overline}

\def\CN{{\cal N}}

\def\rt2{\sqrt{2}}
\def\irt2{{1\over\sqrt{2}}}

\font\cmss=cmss10
\font\cmsss=cmss10 at 7pt

\def\IL{\relax{\rm I\kern-.18em L}}
\def\IH{\relax{\rm I\kern-.18em H}}
\def\rlx{\relax\leavevmode}
\def\ZZ{\rlx\leavevmode\ifmmode\mathchoice{\hbox{\cmss Z\kern-.4em Z}}
{\hbox{\cmss Z\kern-.4em Z}}{\lower.9pt\hbox{\cmsss Z\kern-.36em Z}}
{\lower1.2pt\hbox{\cmsss Z\kern-.36em Z}}\else{\cmss Z\kern-.4em
Z}\fi}
\DeclareMathOperator{\Ai}{Ai}
\DeclareMathOperator{\Tr}{Tr}

\DeclareMathOperator{\Li}{Li}

\DeclareMathOperator{\vol}{vol}
\DeclareMathOperator{\diag}{diag}

\def\bZ {\mathbb{Z}}

\newcommand{\beq}{\begin{equation}}
\newcommand{\eeq}{\end{equation}}
\newcommand{\bal}{\begin{equation}\begin{aligned}}
\newcommand{\eal}{\end{aligned} \end{equation}}
\newcommand{\bea}{\begin{eqnarray}}
\newcommand{\eea}{\end{eqnarray}}

\newcommand{\eqn}[1]{(\ref{#1})}

\newcommand{\cN}{{\mathcal N}}

\newcommand{\cR}{{\mathcal R}}
\newcommand{\cO}{{\mathcal O}}

\newcommand{\CV}{{\mathcal V}}

\title{\boldmath\center{Localization in Supergravity}\\ and \\Quantum $AdS_4/{\rm CFT}_3$ Holography}

\preprint{}
\author{Atish Dabholkar$^{a,b}$, Nadav Drukker$^c$, Jo\~ao Gomes$^d$ \\
\it $^a${Sorbonne Universit\'es, UPMC Paris 06,
UMR 7589, LPTHE, 75005, Paris, France}\\
$^b${CNRS, UMR 7589, LPTHE, 75005, Paris, France}\\
$^c$Department of Mathematics, King's College,
The Strand, WC2R 2LS, London, UK\\
$^d$DAMTP, Center for Mathematical Sciences, University of Cambridge, Wilberforce road, Cambridge, CB3 0WA, UK\\

Emails: \email{atish at lpthe.jussiu.fr}, 
\email{nadav.drukker@gmail.com}, 
\email{jmg84 at cam.ac.uk}
}

\abstract{We compute the quantum gravity partition function of M-theory on $AdS_4 \times X_7 $ by using localization techniques in four-dimensional  gauged supergravity obtained by a consistent truncation on the Sasaki-Einstein manifold $X_{7}$. The supergravity path integral reduces to a finite dimensional integral over two collective coordinates that parametrize the localizing instanton solutions. The renormalized  action of the off-shell instanton solutions depends linearly and holomorphically on  the ``square root'' prepotential  evaluated at the center of $AdS_{4}$. The partition function  resembles the Laplace transform of  the wave function of a topological string and  with an assumption about the measure for the localization integral yields an Airy function in precise agreement with the computation from the boundary ABJM theory on a 3-sphere. Our bulk quantum gravity computation  is nonperturbatively exact in four-dimensional Planck length but ignores corrections due to brane-instantons.}


\begin{document}
\section{Introduction}
\label{sec:Intro}

Holography is one of the fundamental principles of quantum gravity to have emerged from the study of black holes and black branes. A concrete realization of holography is provided within the framework of string theory as an equivalence between quantum gravity in bulk anti de Sitter space and a quantum field theory living at the boundary \cite{Maldacena:1997re}. In its most ambitious formulation this is supposed to be an exact quantum equivalence between the two theories.

While there is substantial evidence for this equivalence, it is far from being fully understood. In particular, a bulk of the earlier work is in the limit of infinite $N$ in the 't~Hooft expansion, which corresponds to \textit{classical} gravity in the bulk. As a result, most applications of holography have been in the direction of using the simpler classical gravitational description to learn about the strongly coupled behavior of large $N$ quantum field theories. 

Given the centrality of the notion of holography for quantum gravity,
it is clearly important to go beyond the classical limit and study quantum effects in the bulk. After all, a primary motivation for string theory is the possibility of unifying general relativity with quantum mechanics. These quantum effects in the bulk correspond to finite $N$ effects in the boundary theory which are of interest in their own right. An important advantage of the $AdS/CFT$ correspondence is that the boundary theory can guide the computation in the bulk. This is especially useful given the notorious difficulties in making sense of the functional integral of quantum gravity.

Such a study of finite $N$ effects has met with considerable success in the context of $AdS_{2}/CFT_{1}$ holography which arises near the horizon of dyonic supersymmetric black holes in string theory. The logarithm of the gravity functional integral gives the quantum entropy of the black hole which is the full quantum generalization of the Bekenstein-Hawking entropy \cite{Sen:2008vm}. The role of $N$ is played by the charges of the black hole. Large charge limit corresponds to the large area limit  for  the black hole horizon. In this limit, the quantum entropy reduces to the Bekenstein-Hawking-Wald entropy. Remarkably, in the simple example of one-eighth BPS black holes in $\CN=8$ supersymmetric theory in four dimensions, it seems possible to evaluate the quantum gravity functional integral exactly, summing all perturbative and nonperturbative corrections to the area formula \cite{Dabholkar:2011ec,Dabholkar:2010uh,Dabholkar:2014ema}. Together, they beautifully reproduce a precise integer\footnote{All essential physical ingredients of the Rademacher expansion  required for obtaining a precise integer can be computed, but  a few technical subtleties remain to be understood better. See \cite{Dabholkar:2014ema} for a discussion.} that equals the quantum degeneracy obtained from the microscopic counting. 

The evaluation of the quantum gravity functional integral 
in $AdS_{2}$ is made possible by the use of localization techniques \cite{Pestun:2007rz,kwy1,Dabholkar:2011ec,Gomes:2013cca} that `localize' the functional integral onto an integral on a finite-dimensional submanifold in field space that parametrizes the `localizing instanton solutions'. In simple cases even this finite-dimensional integral can be evaluated analytically giving a complete solution to the problem. Our goal in this paper is to extend these localization techniques in supergravity to evaluate the bulk quantum gravity partition function in the context of $AdS_{4}/CFT_{3}$ holography.

To describe our results more concretely we recall here a few relevant results from the $AdS_{4}/CFT_{3}$ correspondence. 
This holographic relation is obtained by considering $N$ M2-branes in M-theory on 
$\mathbb{R}^{8}/\mathbb{Z}_{k}$. M-theory on the near horizon $AdS_{4} \times S^{7}/\mathbb{Z}_{k}$ 
geometry is holographically dual to the ABJM Chern-Simons-matter theory \cite{Aharony:2008ug}. The partition function 
of the ABJM theory on $S^{3}$ \cite{fhm,mp-fermi} can be written in terms of the Airy function\footnote{As we review below, a similar expression is valid for a much larger class of 3d quiver 
Chern-Simons-matter theories dual to M-theory on a tri-Sasaki-Einstein manifold $X_7$. While our discussion here focuses on ABJM 
theory, our results apply in the more general setup.} for large $N$ in the 't~Hooft limit (large $N/k$):
\beq
\label{Z-ABJM}
Z^\text{ABJM} \propto\Ai \left[\left(\frac{\pi^2k}{2}\right)^{1/3}\left(N-\frac{k}{24}-\frac{1}{3k}\right)\right] \, .
\eeq
which is \textit{perturbatively exact} but ignores nonperturbative effects at large $N$. 
It is convenient to write the argument of the Airy function as
\beq\label{zdef}
z:= \left(\frac{\pi^2k}{2}\right)^{1/3}\left(N-\frac{k}{24}-\frac{1}{3k}\right) \equiv\left(\frac{\pi^2k}{2}\right)^{1/3} {\hat{N}}
\eeq
The asymptotic expansion of the Airy function for large $z$ is given by
\be\label{asymptotics}
\Ai (z) \sim \frac{1}{2\sqrt{\pi} z^{1/4}} e^{-\frac{2}{3} z^{3/2}} \left( 1 - \frac{5}{48z^{3/2}} + \frac{385}{4608 z^{3}} + \ldots \right) \, , \qquad z \rightarrow \infty \, .
\ee

It is useful to express these results in the Type-IIA string frame. M-theory on $AdS_{4} \times S^{7}/\mathbb{Z}_{k}$ is dual to Type-IIA on $AdS_{4} \times \mathbb{CP}^{3}$ with $N$ units of $F_{4}$ flux through $AdS_{4}$ and $k$ units of $F_{2}$ flux through $\mathbb{CP}^{1} \subset \mathbb{CP}^{3}$ \cite{Aharony:2008ug}. 
Weak coupling and low curvature limit corresponds to the large $N$ and large $k$ limit keeping the radius of curvature $R$ of $AdS_{4}$ large in units of the string length $l$. In this limit, $z$ is large and $N \sim \hat{N}$. In the Type-IIA frame, the 4d string coupling constant $g_{4}$ and the radius $R$ are related to the integers $k$ and $N$ by
\be
g_{4}^{2} \sim \frac{1}{ N k} \sim \frac{1}{N^{2}}\frac{N}{k} \, , \qquad \, \frac{R^{2}}{l^{2}} \sim \left( \frac{{N}}{k} \right)^{\frac{1}{2}}
\ee
The leading behavior of the free energy can be read off from 
\eqref{asymptotics}:
\beq\label{F-ABJM}
F^\text{ABJM} := -\log (Z^{ABJM}) \sim \frac{2}{3} z^{3/2} 
= \frac{{\pi\sqrt{2}}}{3}\, N^{2}\, \left(\frac{k}{N}\right)^{\frac{1}{2}} 
\sim \frac{R^{2}}{g_{4}^{2} l^{2}} \sim \frac{R^{2}}{l_{4}^{2}}
\eeq
where $l_{4} = g_{4} l$ is the 4d Planck length. 
This reproduces the well-known $N^{3/2}$ behavior characteristic of M2-branes which on the gravity side is obtained by evaluating the two-derivative action which equals the square of the $AdS_{4}$ radius in 4d Planck units. 

We see from \eqref{asymptotics} that the expansion parameter is $z^{3/2} \sim {R^{2}}/{l_{4}^{2}}$. The Airy function captures an infinite number of perturbative corrections to the classical expression \eqref{F-ABJM} for the free energy. The logarithmic correction to the free energy was found recently in \cite{Bhattacharyya:2012ye} by computing the one-loop determinants of massless fields around the classical solution. Our objective will be to reproduce the entire Airy function exactly.

An important subtlety concerns the shift in $N$ to $\hat N$. It is convenient to write it as
\be
\frac{\hat{N}}{k} = \frac{N}{k}\left( 1 - \frac{1}{24}\frac{k}{N} - \frac{1}{3}\frac{1}{k N} \right) \sim \frac {N}{k} \left( 1 + C_{1} \frac{l^{4}}{R^{4}} + C_{2} \, g_{4}^{2}\, \right) \, ,
\ee
where $C_{1}$ and $C_{2}$ are some constants. We see that 
$C_{1}$ is a tree level higher derivative contribution likely from the term
$
\int B \wedge I_{8}
$
whereas $C_{2}$ is a one-loop correction. A partial explanation of this shift can be found in \cite{Bergman:2009zh}.
In what follows, we do not attempt to explain this shift but  simply take the radius of $AdS_{4}$ in 4d Planck units as being given, ignoring the distinction between $N$ and $\hat N$. It would be interesting to compute this  shift from a string computation in the bulk.

With this background we can now summarize the main results of the present work.
\begin{myitemize}
\item We obtain a two parameter family of localizing instanton solutions using the off-shell formalism for superconformal gauged supergravity. These solutions with a nontrivial profile for auxiliary fields explore regions in field space faraway from the on-shell solution.
\item Using the square-root prepotential for the truncation of M-theory to $AdS_4$ \cite{Gauntlett:2009zw,Hristov:2009uj} we evaluate the renormalized off-shell action on the localizing submanifold. This off-shell action is evaluated analytically and has a particularly simple form \eqref{ren-action} that depends linearly and holomorphically on  the  prepotential  evaluated at the center of $AdS_{4}$. At its extremum it equals, as expected, the free energy \eqref{F-ABJM} obtained from the on-shell action. 
\item Assuming a simple  measure for the fields we can reproduce the Airy function with the right coefficients as predicted from the gauge theory.  At present we are not able to derive this measure from first principles  because of possible subtleties with gauge fixing to Poincar\'e supergravity and we hope that this question will be resolved in the near future.
\item We emphasize that the localizing solutions are determined entirely by the off-shell supersymmetry transformations and do not depend on the form of the on-shell action. Thus, reduction of the functional integral onto a two-dimensional ordinary integral is already an enormous simplification even without knowing  the physical action and the measure. 
For this reason, the same localizing submanifold could be relevant for evaluating expectation values of the operators  loops that preserve the supersymmetries used for localization.

\item Our results are exact ignoring nonperturbative corrections from world-sheet instantons and Euclidean D2-brane instantons. These nonperturbative  corrections  can in principle be computed by including their contributions  to the Wilsonian effective action of supergravity used for computing the renormalized action on the localizing submanifold\footnote{In principle, there could be additional saddle points as for example the orbifold saddle points that give the nonperturbative corrections to the black hole entropy \cite{Dabholkar:2014ema}.}.

\item In the boundary gauge theory, the Airy function has a natural representation as a Laplace-like symplectic transform of the  tree-level cubic prepotential of   the topological string on local $\mathbb{CP}^{1} \times \mathbb{CP}^{1}$ in the large-radius frame \cite{mp-fermi}.  For the ABJM theory, it has been proposed, using the matrix model representation, that the full non-perturbative completion of the Airy function  is given by  an integral transform of the wave function of the refined topological string on local $\mathbb{CP}^{1} \times \mathbb{CP}^{1}$    \cite{dmp1, dmp2, Hatsuda:2013oxa, Kallen:2013qla}. It is thus tempting to conjecture that this partition function is related to the exponential of the prepotential of the  gauged supergravity truncation on $\mathbb{CP}^{3}$. Note that even though there is no \textit{a priori} notion of a topological string for $\mathbb{CP}^{3}$, the prepotential for the chiral effective action in supergravity is a well-defined object. 

\end{myitemize}

The organization of this paper is as follows. In~\S\ref{sec:MM} we review the localization of the boundary field theory and the evaluation of the partition function in terms of  the Airy function. In~\S\ref{sec:formalism} we review the off-shell $\CN=2$ supergravity formalism and the field content of varies offshell  supermultiplets. In~\S\ref{sec:AdS4} we describe the square-root prepotential relevant to this problem and the supersymmetries of the vacuum $AdS_{4}$ solution. In~\S\ref{sec:Localization} we describe in the bulk supergravity theory. We find the instanton solutions that preserve the localizing supersymmetry and evaluate the renormalized action. With an assumption about the measure, we then show that the bulk partition function is proportional to an Airy function.

\section{Localization in the Boundary Gauge Theory  and the Airy Function}
\label{sec:MM}

We now describe the computation of  the partition function of a large class of models described by a Chern-Simons-Matter gauge theory on the $S^{3}$ boundary of Euclidean $AdS_{4}$. Our goal in the subsequent sections will be to reproduce this partition function from the evaluation of the quantum partition function of the dual gravitational theory.

Supersymmetric localization simplifies the calculation of a partition function of
an arbitrary Chern-Simons-Matter theory in 3d with $\cN=2$ supersymmetry. The 
full path integral on $S^3$ reduces to a finite dimensional integral over a constant mode of a 
single scalar field for each vector multiplets \cite{kwy1,hhl1}.

The resulting integral - a matrix model - can be solved by varied techniques. In simple examples 
the integrals can be done directly in terms of known functions. In the case of ABJM theory, 
the matrix model is very similar to that of the lens space Chern-Simons matrix model for 
which the spectral curve is known \cite{Marino:2002fk, Aganagic:2002wv,Halmagyi:2003ze}. 
This solution was used in \cite{mp-wl} to calculate the 
expectation value of Wilson loops and in \cite{dmp1} to calculate the $S^3$ partition function. 
The free energy exhibits a scaling of $N^{3/2}$ where $N$ is the rank of the gauge group 
with the exact numerical prefactor matching the regularized classical action of 11d SUGRA on 
$AdS_4\times S^7/\bZ_k$. 
Generalizations of this large $N$ behavior were found for a large class of theories and the 
prefactor consistently matched the classical supergravity action of the dual M-theory on 
$AdS_4\times X_7$, which is proportional to the volume of the compact $X_7$ 
\cite{Herzog:2010hf,Santamaria:2010dm,Martelli:2011qj,Cheon:2011vi,Jafferis:2011zi,Gabella:2011sg}.

In the specific case of ABJM theory the solution of the matrix model includs recursive 
corrections in $1/N$ which satisfy a holomorphic anomaly equation \cite{dmp2}. 
It was shown in \cite{fhm} that if nonperturbative effects in M-theory are ignored 
(M2-brane instantons), then this equation is solved by the
Airy function \eqref{Z-ABJM}
These results were rederived directly from the matrix model and generalized to 
other theories with $\cN\geq3$ SUSY in \cite{mp-fermi}, as we review now.

The theories we consider are neckless quivers with $r$ copies of $U(N)$ gauge groups 
labeled by $a=1,\cdots,r$. Each gauge group  has a Chern-Simons term at level 
$k^{(a)}=n^{(a)}k$ with $\sum n^{(a)}=0$,  bifundamental chiral multiplets, and in addition an 
arbitrary number $N^{(a)}_f$ of chiral 
multiplets in the fundamentals of the gauge groups. In this case the matrix model of \cite{kwy1} is
\beq
\label{gen-quiv}
Z(N)=\frac{1}{(N!)^r}
\int\prod_{a,i}\frac{d\mu^{(a)}_i}{2\pi}
\frac{\exp\left[\frac{in^{(a)}k}{4\pi}\big(\mu^{(a)}_i\big)^2\right]}
{\Big(2\cosh\frac{\mu^{(a)}_i}{2}\Big)^{N^{(a)}_f}}
\prod_{a=1}^r\frac{\prod_{i<j}\Big(2\sinh\frac{\mu^{(a)}_i-\mu^{(a)}_j}{2}\Big)^2}
{\prod_{i,j}2\cosh\frac{\mu^{(a)}_i-\mu^{(a+1)}_j}{2}}
\eeq

The particular case of ABJM theory is when $r=2$, $n_1=-n_2=1$ and there is no fundamental matter, so 
\beq
Z_\text{ABJM}(N)=\frac{1}{(N!)^2}
\int\prod_{i}\frac{d\mu_id\nu_i}{(2\pi)^2}
\exp\left[\frac{ik}{4\pi}\big(\mu_i^2-\nu_i^2\big)\right]
\frac{\prod_{i<j}\Big(2\sinh\frac{\mu_i-\mu_j}{2}\Big)^2\Big(2\sinh\frac{\nu_i-\nu_j}{2}\Big)^2}
{\prod_{i,j}\big(2\cosh\frac{\mu_i-\nu_j}{2}\big)^2}
\eeq

The main idea of \cite{mp-fermi} was to  first compute a grand canonical potential instead of computing the canonical partition function directly:
\beq
\Xi(z)=1+\sum_{N=1}^\infty z^NZ(N)
\eeq
Let us also define $\mu=\log z$ and $J(\mu)=\log\Xi(e^\mu)$. 

By use of the Cauchy identity one can rewrite (\ref{gen-quiv}) as
\bal
Z(N)=\frac{1}{N!}
\sum_{\sigma\in S_N}(-1)^\sigma&\int\prod_i\frac{d\mu^{(1)}_i}{2\pi}
\rho(\mu^{(1)}_i,\mu^{(1)}_{\sigma(i)})\,,
\\
\rho(\mu^{(1)}_i,\mu^{(1)}_j)
&=\int\prod_{a=2}^{r}\frac{d\mu^{(a)}_j}{2\pi}
K_1\big(\mu^{(1)}_i,\mu^{(2)}_j\big)
\prod_{a=2}^{r}K_{a}\big(\mu^{(a)}_j,\mu^{(a+1)}_j\big)\,,
\\
K_a\big(\mu,\mu'\big)=&
\frac{\exp\left[\frac{in^{(a)}k}{4\pi}\mu^2\right]}
{\Big(2\cosh\frac{\mu}{2}\Big)^{N^{(a)}_f}}
\frac{1}{2\cosh\frac{\mu-\mu'}{2}}
\eal
We can view $\rho(\mu,\nu)$ as the matrix elements of a density matrix $\hat\rho=e^{-\hat H}$ and then the product over $\rho$ can be rearranged into a sum over conjugacy classes of the permutation group ({\em i.e.} partitions of $N$) of $\Tr (\hat\rho)^{l_\alpha}$ with the appropriate multiplicities. The analogous expression for the grand canonical partition function is much simpler than for the canonical one and is given by a Fredholm determinant
\beq
\label{Xi}
\Xi=\det(1+z\hat\rho)=\prod_n(1+e^{\mu-E_n})
\eeq
where $E_n$ are the eigenvalues of the Hamiltonian $\hat H$.

It can be shown \cite{mp-fermi} that classically this hamiltonian is
\beq
H(p,q)=\log\left[2\cosh\frac{q}{2k}\right]\sum_{a=1}^rN_f^{(a)}
+\sum_{a=1}^r\log\left[2\cosh\frac{p-\sum_{b=1}^an^{(b)}q}{2}\right]
\eeq
For large $p$ and $q$ this is simply
\beq
\label{poly}
H(p,q)\approx\frac{1}{2}\left(|q|\sum_{a=1}^r\frac{N_f^{(a)}}{k}
+\sum_{a=1}^r\left|p-\sum_{b=1}^an^{(b)}q\right|\right)
\eeq

In the thermodynamic limit, the density of eigenvalues of $\hat H$ is the derivative of the number of eigenstates below a given energy
\beq
\rho(E)=\sum_n\delta(E-\lambda_n)=\frac{dn(E)}{dE}
\eeq
In our case $n(E)$ is given by the volume of phase space bound by the polygon $H=E$ in the $(q,p)$ plane \eqn{poly}, which is of the form $n(E)=\frac{c}{k}E^2$, where $c$ is a constant which depends on the numbers $N^{(a)}_f$ and $n^{(b)}$ and is easily determined from \eqn{poly}. The inclusion of the first quantum corrections shifts this by a constant $n_0$ at large $E$. The expected behavior at large and small $E$ is captured by
\beq
\label{nE}
n(E)=\frac{c}{k}E^2+n_0(1-e^{-E})
\eeq
The full argumentation as well as the derivation of $n_0$ for some examples can be found in \cite{mp-fermi}.

The one particle partition function is given by
\beq
\Tr(\hat\rho^l)=\int_0^\infty dE\, \rho(E)e^{-lE}
\eeq
and the grand canonical potential is \eqn{Xi}
\beq
J(\mu)=\int_0^\infty dE\,\rho(E)\log(1+e^{\mu-E})
=-\frac{2 c}{k}\Li_3(-e^{\mu })+n_0\mu (1+e^{-\mu }) \log (1+e^{-\mu })-n_0
\eeq
For large $\mu$ this is
\beq\label{grand canonical pot}
J(\mu)\approx\frac{c}{3k}\mu^3+\left(\frac{\pi^2c}{3k}+n_0\right)\mu-A
\eeq
The first two terms depend only on the quadratic growth in the number of states and the constant shift $n_0$. The ansatz for the non-perturbative corrections in \eqn{nE} gives $A=n_0$, but it is possible to modify the $e^{-E}$ term without changing the large $E$ or $E\sim0$ behavior and this will affect the constant term, so we leave $A$ as an undetermined constant.

Now the canonical partition function can be derived from the grand canonical potential by
\beq
\label{airy}
Z(N)=\frac{1}{2\pi i}\int d\mu\,e^{J(\mu)-N\mu}
=\left(\frac{c}{k}\right)^{-1/3}e^A\Ai\left[\left(\frac{c}{k}\right)^{-1/3}\left(N-\frac{\pi^2c}{3k}-n_0\right)\right]\,,
\eeq
In the case of ABJM theory the constants are $c=2/\pi^2$ and $n_0=k/24-1/3k$, which indeed 
reproduces \eqn{Z-ABJM}

This Airy function expression is corrected by non-perturbative $O(e^{-N})$ terms, which we will not study. The three parameters: $c$, $n_0$ and $A$ depend only on the $\hbar=1/2\pi k$ expansion of the one-particle hamiltonian. $c$ depends only on the classical Hamiltonian and is therefore exact. $n_0$ depends on the $O(\hbar)$ terms in the Hamiltonian as well as the quantum corrections to the phase space boundary and is expected to be at most linear in $k$. The parameter $A$ may have higher order dependence in $k$. Even though the expansion is around $k=0$, the fact that it is perturbatively exact in $N$ means that the argument of the Airy function is a robust expression valid both in the 't~Hooft limit (large $N$ fixed $N/k$) and the M-theory regime (large $N$ fixed $k$). 

\section{Superconformal Formalism and Gauged Supergravity}
\label{sec:formalism}

We work with the action in four dimensions with $\CN=2$ supersymmetry obtained 
as a consistent truncation of M-theory action on a Sasaki-Einstein manifold $X_{7}$.
As we will describe in~\S\ref{sec:AdS4}, the nearly massless fields that we work with consist of 
the supergravity multiplet, a universal hypermultiplet (or the dualized tensor multiplet), and a single vector multiplet\footnote{We require the Euclidean continuation of this theory. For the purposes of localization  we relax the reality conditions on the fields and look for complex saddle points in field space.}. 

For the purposes of localization it is most convenient to use an off-shell formalism to describe the supersymmetry variations of all these multiplets. The main advantage is that the off-shell supersymmetry transformations do not depend on the form of the physical on-shell action. As a result the localizing instanton solution determined entirely by the supersymmetry transformations is universal and therefore valid for any on-shell action. 

Ordinary supergravity theories are invariant under local coordinate transformations, local lorentz transformations and supertranslations. However an off-shell realization of the super-Poincare algebra is usually very difficult, if at all possible. A different approach consists of enlarging the super-Poincare group by including local superconformal transformations and then using gauge fixing to reduce the theory to the usual super-Poincare gravity. The great advantage of this formalism is that an off-shell construction is more easily implemented. In the following we review the superconformal multiplet field contents and SUSY transformations.

The relevant $\mathcal{N}=2$ superconformal algebra consists of the generators,
\beq
P_a,
\quad
Q_i,
\quad
M_{ab},
\quad
D,
\quad
U_{ij},
\quad
S^i,
\quad
K_a
\eeq
where $P_a,\,M_{ab},\,D$ and $K_a$ are respectively the generators of translations, Lorentz rotations, dilatations and conformal boosts. They close into the conformal algebra. The remaining generators $Q_i$ and $S^i$ encode usual supersymmetry and special conformal supersymmetry respectively. The generator $U_{ij}$ corresponds to the four dimensional $SU(2)$ R-symmetry.

In the superconformal formalism we add the gauge fields corresponding to the generators above
\beq
e^a_{\mu},
\quad
\psi^i_{\mu},
\quad
\omega_{\mu}^{ab},
\quad
b_{\mu},
\quad
\mathcal{V}^{ij}_{\mu},
\quad
\phi^i_{\mu},
\quad
f^a_{\mu}
\eeq
respectively. Here $a$ and $\mu$ denote tangent and coordinate space indices respectively. Due to a large gauge freedom we can impose additional constraints on the gauge fields via their curvatures. This allows us to express the $M,\, S$ and $K$ gauge fields, respectively $\omega^{ab}_{\mu}$, $\phi^i_{\mu}$ and $f^a_{\mu}$, in terms of the other fields. After this we can identify $e^a_{\mu}$ and $\omega^{ab}_{\mu}$ as the vielbein and spin connection respectively. Additional gauge fixing conditions can be used to fix both conformal invariance and $SU(2)$ R-symmetry leaving behind a super-Poincare subgroup. 

In the following sections we present the field contents of the various supermultiplets of superconformal supergravity and respective off-shell actions. We follow closely the four dimensional construction of \cite{deWit:1984px}.

\subsection{Weyl Multiplet}
The Weyl multiplet, which we denote here by $\bf W$, contains 24+24 independent field components
\be\label{Weylfields}
{\bf W} = \left( e_{\mu}^{a}, \psi_{\mu}^{i}, b_{\mu}, A_{\mu}, \CV_{\mu \, j}^{\, i}, T_{ab}^{ij}, \chi^{i}, D \right),
\ee
$e_{\mu}^{a}$ is the vielbien, 
$\psi_{\mu}^{i}$ is the (left-handed) gravitino doublet associated with the $N=2$ supersymmetries, 
$b_{\mu}$ and $A_{\mu}$ are the gauge fields of dilatations and chiral $U(1)$ R-symmetry transformations respectively, and 
$\CV_{\mu \, j}^{\, i}$ are the gauge fields for the $SU(2)$ R-symmetries. 
The $SU(2)$ doublet of Majorana spinors $\chi^{i}$, the antisymmetric anti-selfdual  field $T_{ab}^{ij}$ 
and the real scalar field $D$ are all auxiliary fields. 

Under the Q and S supersymmetry and special conformal
transformations, with parameters $\epsilon, \eta,\Lambda^a_K $ respectively, the independent fields of the Weyl multiplet transform
as:
\begin{align}
\label{eq:weyl-multiplet}
\delta e_\mu{}^a =&\, \bar{\epsilon}^i \, \gamma^a \psi_{ \mu i} +
\bar{\epsilon}_i \, \gamma^a \psi_{ \mu}{}^i \, , \nonumber\\
\delta \psi_{\mu}{}^{i} =&\, 2 \,\mathcal{D}_\mu \epsilon^i - \tfrac{1}{8}
T_{ab}{}^{ij} \gamma^{ab}\gamma_\mu \epsilon_j - \gamma_\mu \eta^i
\, \nonumber \\
\delta b_\mu =&\, \tfrac{1}{2} \bar{\epsilon}^i \phi_{\mu i} -
\tfrac{3}{4} \bar{\epsilon}^i \gamma_\mu \chi_i - \tfrac{1}{2}
\bar{\eta}^i \psi_{\mu i} + \mbox{h.c.} + \Lambda^a_K e_{\mu a} \, ,
\nonumber \\
\delta A_{\mu} =&\, \tfrac{1}{2} i \bar{\epsilon}^i \phi_{\mu i} +
\tfrac{3}{4} i \bar{\epsilon}^i \gamma_\mu \, \chi_i +
\tfrac{1}{2} i\bar{\eta}^i \psi_{\mu i} + \mbox{h.c.} \, , \nonumber\\
\delta \mathcal{V}_\mu{}^{i}{}_j =&\, 2\, \bar{\epsilon}_j
\phi_\mu{}^i - 3
\bar{\epsilon}_j \gamma_\mu \, \chi^i + 2 \bar{\eta}_j \, \psi_{\mu}{}^i
- (\mbox{h.c. ; traceless}) \, , \nonumber \\
\delta T_{ab}{}^{ij} =&\, 8 \,\bar{\epsilon}^{[i} R(Q)_{ab}{}^{j]} \,
, \nonumber \\
\delta \chi^i =&\, - \tfrac{1}{12} \gamma^{ab} \, \slashed{D} T_{ab}{}^{ij}
\, \epsilon_j + \tfrac{1}{6} R(\mathcal{V})_{\mu\nu}{}^i{}_j
\gamma^{\mu\nu} \epsilon^j -
\tfrac{1}{3} i R_{\mu\nu}(A) \gamma^{\mu\nu} \epsilon^i + D
\epsilon^i +
\tfrac{1}{12} \gamma_{ab} T^{ab ij} \eta_j \, , \nonumber \\
\delta D =&\, \bar{\epsilon}^i \, \slashed{D} \chi_i +
\bar{\epsilon}_i \,\slashed{D}\chi^i \, .
\end{align} where we defined the covariant derivative as follows
\begin{equation}
\label{eq:D-epslon}
\mathcal{D}_{\mu} \epsilon^i = \big(\partial_\mu + \tfrac{1}{4}
\omega_\mu{}^{cd} \, \gamma_{cd} + \tfrac1{2} \, b_\mu +
\tfrac{i}{2} \, A_\mu \big) \epsilon^i + \tfrac1{2} \,
\mathcal{V}_{\mu}{}^i{}_j \, \epsilon^j \,.
\end{equation}

\subsection{Vector Multiplet}
The field content of the vector multiplet consists of
\be\label{Vectorfields}
{\bf X}^{I} = \left( X^{I}, \O_{i}^{I}, W_{\mu}^{I}, Y^{I}_{ij} \right)
\ee
with $8+8$ degrees of freedom. $X^{I}$ is a complex scalar, the gaugini $\O^{I}_{i}$ are the $SU(2)$ 
doublet of chiral fermions, $W^{I}_{\mu}$ is a vector field, and $Y^{I}_{ij}$ are an $SU(2)$ triplet of 
auxiliary scalars (this means $Y_{ij}=Y_{ji}$ and $Y_{ij}=\epsilon_{ik}\epsilon_{jl}Y^{kl}$). The index $I$ denotes the generators $t_I$ of the gauge group $G$.

The Q, S-transformations of the vector multiplet fields are as follows
\begin{align}
\label{eq:variations-vect-mult}
\delta X =&\, \bar{\epsilon}^i\Omega_i \,,\nonumber\\
\delta\Omega_i =&\, 2 \slashed{D} X\epsilon_i
+\frac{1}{2} \varepsilon_{ij} \mathcal{F}_{\mu\nu}^-
\gamma^{\mu\nu}\epsilon^j +Y_{ij} \epsilon^j
+2X\eta_i\,,\nonumber\\
\delta W_{\mu} = &\, \varepsilon^{ij} \bar{\epsilon}_i
(\gamma_{\mu} \Omega_j+2\,\psi_{\mu j} X)
+ \varepsilon_{ij}
\bar{\epsilon}^i (\gamma_{\mu} \Omega^{j} +2\,\psi_\mu{}^j
\bar X)\,,\nonumber\\
\delta Y_{ij} = &\, 2\, \bar{\epsilon}_{(i}
\slashed{D}\Omega_{j)} + 2\, \varepsilon_{ik}
\varepsilon_{jl}\, \bar{\epsilon}^{(k} \slashed{D}\Omega^{l)
} \,,
\end{align}
and we defined $\mathcal{F}_{\mu\nu}=F_{\mu\nu}-1/4\left(\bar{X}\epsilon_{ij}T^{ij}_{\mu\nu}+h.c.\right)+\text{fermions}$ with $F_{\mu\nu}=\partial_{\mu}W_{\nu}-\partial_{\nu}W_{\mu}$.

For the problem at hand we will take the gauge group to be $G=U(1)^{n_V+1}$ where $n_v$ is the number of physical vector multiplets. Note that the Weyl multiplet does not have any physical vector and therefore we need to add a compensating $I=0$ vector multiplet.

\subsection{Hypermultiplet}\label{Hypermultiplet sec}

It is well-known that for the hypermultiplets of $\CN=2$ supersymmetry, off-shell closure of the supersymmetry algebra cannot be achieved  with finite number of fields. We first describe the on-shell hypermultiplet and then describe how off-shell closure can be achieved by adding an infinite sequence of fields. 

The hypermultiplets are based on scalars $A_i^{\;\alpha}$ and spinors $\zeta^{\alpha}$. The scalars are doublets under the $SU(2)$ four dimensional R-symmetry and transform in the fundamental of $Sp(2r)$ so that the index $\alpha$ takes values in $1\ldots 2r$. The scalar fields satisfy a reality constraint
\begin{equation}
\label{reality A}
A^i_{\;\alpha}=(A_i^{\;\alpha})^{*}=\epsilon^{ij}\rho_{\alpha\beta}A_j^{\;\beta}
\end{equation}
from which follows the consistency condition $\rho_{\alpha\beta}\rho^{\beta\gamma}=-\delta^{\gamma}_{\alpha}$, with $\rho_{\alpha\beta}$ a $2r\times 2r$ matrix. The scalars can be seen as sections of a $4r$ hyperkahler manifold. In addition a subgroup $G'$ of the gauge group $G$ can act on the index $\alpha$ as a subgroup of $Sp(2r)$. We will see later on that it is the action of this subgroup $G'$ on the hypers that generates a cosmological constant.

The transformations under Q- and S-supersymmetry, with parameters $\epsilon^i$ and $\eta_i$ respectively, are
\begin{eqnarray}
&&\delta A_{i}^{\;\alpha}=2\bar{\varepsilon}_i\zeta^{\alpha}+2\rho^{\alpha\beta}\varepsilon_{ij}\bar{\varepsilon}^{j}\zeta_{\beta}\\
&& \delta \zeta^{\alpha}=\slashed D A_i^{\;\alpha}\varepsilon^i+2gX^{\alpha}_{\;\beta}A_i^{\;\beta}\epsilon^{ij}\varepsilon_j+A_{i}^{\;\alpha}\eta^i\label{SUSY hyper}
\end{eqnarray}where we have defined the Lie-algebra valued quantities
\begin{equation}
\label{X-rep}
X^{\alpha}_{\;\beta}\equiv X^I t_{I\beta}^{\alpha},\;\;\bar{X}^{\alpha}_{\;\beta}\equiv \bar{X}^I t_{I\beta}^{\alpha}
\end{equation}with the generators satisfying the reality condition $t_{\alpha}^{\;\beta}\rho_{\beta\gamma}=\rho_{\alpha\beta}t^{\beta}_{\;\gamma}$. The covariant derivative has the form
\begin{equation}\label{cov deriv hyper}
D_{\mu}A_i^{\;\;\alpha}=\partial_{\mu}A_{i}^{\;\;\alpha}+\frac{1}{2}V_{\mu i}^{j}A_j^{\;\;\alpha}-b_{\mu}A_i^{\;\;\alpha}-gW_{\mu\beta}^{\alpha}A_i^{\;\;\beta}-\bar{\psi}_{\mu i}\zeta^{\alpha}-\rho^{\alpha\beta}\varepsilon_{ij}\bar{\psi}_{\mu}^j\zeta_{\beta}
\end{equation}with $g$ the coupling constant. 

An on-shell counting shows that $A_i^{\;\alpha}$ and $\zeta^{\alpha}$ cannot constitute an off-shell supermultiplet, 
as there are $4r$ bosonic and $8r$ fermionic degrees of freedom. As a matter of fact the superconformal algebra only 
closes provided the fields satisfy additional constraints. These constraints are complete in the sense that they are
invariant under supersymmetry and do not generate further constraints. Further details can be found in \cite{deWit:1984px}. 

To take the hypermultiplet off-shell we need to consider an infinite tower of hypermultiplet fields in 
sequences $(A_i^{\;\alpha},\zeta^{\alpha})$, $(A_i^{\;\alpha},\zeta^{\alpha})^{(z)}$, $(A_i^{\;\alpha},\zeta^{\alpha})^{(zz)}$, etc. All symmetries act identically within each set of $2r$ fields, except for one abelian generator under which $(A_i^{\;\alpha},\zeta^{\alpha})$ goes into $(A_i^{\;\alpha},\zeta^{\alpha})^{(z)}$, $(A_i^{\;\alpha},\zeta^{\alpha})^{(z)}$ into $(A_i^{\;\alpha},\zeta^{\alpha})^{(zz)}$ and so on. This is the central charge generator. Closure of the superconformal algebra implies therefore an infinite set of constraints from which only the fields $(A_i^{\;\alpha},\zeta^{\alpha},A_i^{\;\alpha(z)})$ become independent. This is possible due to the structure of the constraints which consist of Klein-Gordon and Dirac type of equations. Under this setup the SUSY transformation (\ref{SUSY hyper}) gets modified to
\begin{equation}
\delta \zeta^{\alpha}=\slashed D A_i^{\;\alpha}\varepsilon^i+2gX^{\alpha}_{\;\beta}A_i^{\;\beta}\epsilon^{ij}\varepsilon_j+A_{i}^{\;\alpha}\eta^i+aA_{i}^{\;\alpha(z)}\epsilon^{ij}\varepsilon_j
\end{equation}where, as we will see next, the field $A_{i}^{\;\alpha(z)}$ becomes an auxiliary field in the sense that the action does not contain a kinetic term for it. The field $a$ is the scalar field of the vector multiplet associated with the central charge translation and can be gauged away by setting $a=1$. It will not play any role in the following discussion except to produce a mass term for $A_{i}^{\;\alpha(z)}$ in the Lagrangian.

To construct a Lagrangian for the independent fields $(A_i^{\;\alpha},\zeta^{\alpha},A_i^{\;\alpha(z)})$ we need to first  construct a linear multiplet that couples the hypers $(A_i^{\;\alpha},\zeta^{\alpha})$ to $(A_i^{\;\alpha},\zeta^{\alpha})^{(z)}$. This linear multiplet is then contracted with the components of the vector multiplet associated with the central charge resulting in a superconformal invariant density. After using the supersymmetry constraints, which depend on the other hyper families, the resulting Lagrangian ends up depending only on $(A_i^{\;\alpha},\zeta^{\alpha},A_i^{\;\alpha(z)})$ as required. The final expression is \cite{deWit:1984px}
\bal
\label{hypers}
\mathcal{L}&=[-D_{\mu}A^i_{\;\beta}D^{\mu}A_i^{\;\alpha}-\frac{1}{6}RA^i_{\;\beta}A_i^{\;\alpha}+\frac{1}{2}DA^i_{\;\beta}A_i^{\;\alpha}+(|a|^2+W_{\mu}^zW^{\mu z})A^{i\;(z)}_{\;\beta}A_i^{\;\alpha(z)}\\
&\quad{}+4g^2A^i_{\;\beta}\bar{X}^{\alpha}_{\;\gamma}X^{\gamma}_{\;\delta}A_i^{\;\delta}+gA^i_{\;\beta}Y^{jk\,\alpha}_{\gamma}A_k^{\;\gamma}\epsilon_{ij}]d_{\alpha}^{\;\beta}+\text{fermionic terms}
\eal
where $d_\alpha^{\;\beta}$ is a matrix with the follwoing properties:
\bal
\text{Hermitician:}\quad&
\bar{d_\alpha^{\;\beta}}\equiv d_\beta^{\;\alpha}\,,\\
\text{quaternionic:}\quad&
d_\alpha^{\;\,\beta}=\epsilon_{\gamma\alpha}\epsilon^{\delta\beta}d^{\gamma}_{\;\,\delta}\,,\\
\text{gauge invariant:}\quad&
t^{\gamma}_{\;\,\alpha}d_{\gamma}^{\;\,\beta}+d_{\alpha}^{\;\,\gamma}t_{\gamma}^{\;\,\beta}=0\,.
\eal
As shown in appendix of  \cite{deWit:1984px} the matrix $d_\alpha^{\;\beta}$ can be brought to the diagonal form which we choose to be $d_\alpha^{\;\beta}=-\delta^{\beta}_{\alpha}$.

\section{$\mathcal{N}=2$ Supersymmetry and the $AdS_4$ Vacuum Solution}
\label{sec:AdS4}

All  gauge theories described in section~\ref{sec:MM} have a dual description 
as M-theory on $AdS_4\times X_7$, where $X_7$ is an appropriate compact tri-Sasaki-Einstein seven-manifold\footnote{A tri-Sasaki-Einstein manifold is dual to gauge theories with three supersymmetries which exhibit the universal Airy function behavior.}
as described in detail in \cite{Gauntlett:2009zw}.  For the ABJM theory, $X_7=S^7/\mathbb{Z}_k$ which can be seen as an Hopf fibration over $M_6=\mathbb{CP}^{3}$, but our discussion below is applicable to a general $X_7$.%
\footnote{Some examples of $M_6$ for other Sasaki-Einstein Manifolds can be found \textit{e.g.}, in 
\cite{Martelli:2008rt, Martelli:2009ga}.}

Since we do not know how to implement localization directly in the 11-dimensional M-theory, we will use the 4-dimensional
truncation  to $AdS_4$ and restrict to a minimal set of fields which form 
a consistent truncation\footnote{In principle one should include the infinite tower of massive 
Kaluza-Klein modes. We assume that the truncated theory is adequate for discussing localization.}. The truncation has a natural interpretation as a flux compactification on a
6d manifold with $SU(3)$ structure \cite{Gauntlett:2009zw} and the massless sector is easiest to understand in this description. The four scalar fields of the universal hypermultiplet arise from the dilaton, the dualized NS two-form B and a complex scalar arising from the 3-form RR potential $C^{(3)}$ parallel to the (3,0) form on $M_6$. The vector multiplet contains a vector coming from the $C^{(3)}$ field parallel to the K\"ahler form and a complex scalar that corresponds to the complexified K\"ahler modulus. The presence of fluxes and the fact that the (3,0) form is not closed leads to a gauging of the four-dimensional supergravity and gives masses to some of the fields. This truncation is consistent and hence there is no ambiguity about which massive KK modes should be kept in the classical four-dimensional theory.

The resulting truncated theory in four dimensions is described by \cite{Gauntlett:2009zw} 
with the prepotential 
\beq\label{Ourprepot}
F=\sqrt{X^0(X^{1})^3} \, .
\eeq
The reduction of the Sasaki-Einstein manifold naturally leads to a cubic prepotential as for the usual Calabi-Yau reductions. After dualization one obtains the somewhat unusual, non-polynomial, square-root form of the prepotential. The main advantage of the square-root prepotential is that one does not have to deal with a tensor multiplet and the full off-shell supersymmetry transformations in the gauged supergravity are explicitly known. 

To construct an $AdS_{4}$ background we require gauged supergravity, which in this context can be obtained by introducing a charged compensator. The idea is to consider a system of $n_V+1$ vector multiplets coupled to a charged conformal hypermultiplet $A_i^{\;\;\alpha}$, with $\alpha=1,2$. We consider a model with charges \be\label{charged hyper}
t_IA_i^{\;\;\alpha}=P_I(i\sigma^3)_{\;\;\beta}^{\alpha}A_i^{\;\;\beta}
\ee 
where  $P_I$ are the moment maps on the hyperkahler manifold. After fixing the $U$ gauge transformation by $A_i^{\;\;\alpha}\propto \delta_i^{\alpha}$  then generates a negative cosmological constant via the hypermultiplet couplings as we show later on.

In the following sections we describe the $AdS_4$ vacuum of the theory, which preserves eight supercharges, by solving the off-shell SUSY equations in this background. We first solve the gravitini equations to find the Killing spinors and then proceed to find the values of the fields for both the vectors and hypers of the off-shell gauged supergravity. We end by showing that the solutions are consistent with the equations of motion.

\subsection{SUSY Equations}

For the vacuum $AdS_4$ solution  we take the hyperbolic metric
\beq\label{AdS4 metric}
ds^2=L^2\left(d\eta^2+\sinh^2(\eta)d\Omega_3^2\right)
\eeq
where $d\Omega_3^2$ denotes the metric of the round $S^3$, and $L$ is the size of $AdS_4$ in the M-theory frame. We assume that any field with non-zero spin is zero. This is justified by the fact that the geometry does not have any non-trivial cycles. 

We start by solving the BPS equations for the Weyl multiplet. The equation $\delta\chi=0$ gives automatically $D=0$. Note that this field cannot be determined solely by its equation of motion since it acts as a Lagrange multiplier. 
The vanishing of  the gravitini variation gives the Killing spinor equations
\beq 
\delta\psi^i_{\mu}=2\nabla_{\mu}\epsilon^i-\gamma_{\mu}\eta^i=0
\eeq
where $i,j$ are the four dimensional $SU(2)$ R-symmetry indices and we have taken the on-shell values for $b_{\mu},A_{\mu},\mathcal{V}_{\mu}{}^i{}_j=0$ in the covariant derivative (\ref{eq:D-epslon}). We have used the convention that up (down) indices for $\epsilon$ denote positive (negative) chirality and the opposite for $\eta$. The $\delta\psi_i$ equation is obtained by lowering or raising the indices.  In the Euclidean theory, it is more convenient to use the  Dirac notation. For this purpose,  we first relabel different indices as \cite{Dabholkar:2010uh,Banerjee:2011ts}
\beq
\xi^i_+\equiv \epsilon^i,\qquad \epsilon_{i}\equiv i\epsilon_{ij}\xi^j_- ;
\qquad
\eta_{i}\equiv-\epsilon_{ij}\eta^{j}_{+},\qquad\eta^{i}\equiv i\eta_{-}^i
\eeq
and then reassemble these components into Dirac spinors $\xi=(\xi^i_+,\xi^i_-)$ and $\eta=(\eta^i_+,\eta^i_-)$. In the Euclidean theory the two chiral representations are not complex conjugates of each other.

In the Dirac notation, the Killing spinor equation  becomes
\beq\label{Killing eq0}
\delta\psi^i_{\mu}=2\nabla_{\mu}\xi^i-i\gamma_{\mu}\eta^i=0 \, .
\eeq
The $\mathcal{N}=2$ Killing spinor equation in  $AdS_4$  space has the form
\beq\label{AdS4 Killing sp}
\nabla_{\mu}\xi^i=\frac{i}{2L}\gamma_{5}\gamma_{\mu}q^{i}_{\;j}\xi^j
\eeq
with $\mathbf{q}^2=\mathbb{I}$, so that the space has constant negative curvature.
This suggests that we choose $\eta$    in \eqref{Killing eq0} to be of the form \cite{Hanaki:2006pj}
\beq\label{gauge choice}
\eta^i=-\frac{1}{L}\gamma_5\sigma^i_{3j}\xi^j
\eeq
with $\sigma_3$ the Pauli matrix $\diag(1,-1)$.  Equation (\ref{Killing eq0}) now becomes
\beq\label{Killing eq2}
\nabla_\mu \xi^i=\frac{i}{2L}\gamma_5\gamma_{\mu}\sigma^i_{3j}\xi^j
\eeq
in agreement with (\ref{AdS4 Killing sp}). The resulting $Q$-supersymmetry is then a combination of a $Q(\xi)$ and $S(\xi)$-supersymmetries. Explicit solutions of this equation are described in the appendix~\ref{Killing}. 

We  now consider the vector multiplet BPS equations. After setting $F_{\mu\nu}=T_{\mu\nu}=0$ we find 
\bal
\delta \Omega^i_+&=-i\slashed\partial X\xi^i_- -\frac{1}{2}Y^i_{\;j}\xi^j_+ +X\eta^i_+=0\\
\delta \Omega^i_-&=-i\slashed\partial \bar{X}\xi^i_+ -\frac{1}{2}Y^i_{\;j}\xi^j_- +\bar{X}\eta^i_-=0
\eal
where the $\pm$ indices denote chirality. We have denoted $Y^i_{\;\,j}\equiv\epsilon_{jk}Y^{ik}$ so that we also have $Y_{ij}=\epsilon_{ik}Y^{k}_{\;\,j}$. If we parametrize the scalars by $X=H+iJ$ then, in Dirac notation, we have
\beq
\label{vector-BPS}
-i\slashed \partial(H-i\gamma_5 J)\xi^i-\frac{1}{2}Y^i_{\;j}\xi^j-\frac{1}{L}(H+i\gamma_5 J)\gamma_5(\sigma_3)^i_{\;j}\xi^j=0,
\eeq
where we used the choice (\ref{gauge choice}) for $\eta$. This equation must be satisfied for all the eight Killing spinors. It is easy to see that these equations are solved for constant values of the scalars $X,\,Y$. In  appendix~\ref{Loc appendix} we show that this is indeed the only solution. To see that we consider the solution to the BPS equation for a particular Killing spinor, which has a non-trivial spatial dependence (see the next section). Then we notice that when choosing different Killing spinors the solutions are compatible only when the scalars are constant. We therefore have the solution
\beq
H=0,\qquad Y^1_{\;\; 1}=-Y^2_{\;\;2}=-2i \frac{J}{L}\,,
\qquad
Y^1_{\;\;2}=-Y^2_{\;\;1}=0\,.
\eeq
It is simple to check that this is also a solution of the equations of motion (\ref{EOM}), which 
is of course a consequence of supersymmetry.
The value of $J$ is determined by the hypers SUSY equations as we show in the following.

The BPS equations for  the hypers imply
\bal
\delta \zeta_{\alpha +}&=i\slashed \nabla A^i_{\;\; \alpha}\epsilon_{ij}\xi^j_- +2g \bar{X}_{\alpha}^{\;\;\beta}A^i_{\;\;\beta}\epsilon_{ij}\xi^j_+ -A^i_{\;\; \alpha}\epsilon_{ij}\eta^j_+ +\bar{a}A^{i(z)}_{\alpha}\epsilon_{ij}\xi^j_{+}=0\\
\delta \zeta^{\alpha}_-&=\slashed \nabla A_i^{\;\; \alpha}\xi^i_+ +2g i X^{\alpha}_{\;\;\beta}A_i^{\;\;\beta}\epsilon^{ij}\epsilon_{jk}\xi^k_- +iA_i^{\;\; \alpha}\eta^i_- -iaA_i^{(z)\alpha}\xi^i_-=0
\eal
where we have defined $\nabla A^i_{\;\; \alpha}$ by setting the gauge fields and the fermions to zero in (\ref{cov deriv hyper}).
Using the relations \eqn{reality A}, \eqn{X-rep} 
we obtain the SUSY equation in the Dirac basis
\beq
\slashed \nabla A_i^{\;\; \alpha}\xi^i-2g i (H^I-i\gamma_5 J^I)t^{\alpha}_{I\beta}A_i^{\;\; \beta}\xi^i+iA_i^{\;\; \alpha}\eta^i-iF_i^{\;\;\alpha}\xi^i=0
\eeq
where $F_i^{\;\;\alpha}\equiv aA_i^{(z)\alpha}$ is the auxiliary field which has the reality constraint $F_i^{\;\;\alpha}=(F^i_{\;\;\alpha})^*=\epsilon_{ij}\epsilon^{\alpha\beta}F^j_{\;\;\beta}$ \cite{deWit:1984px,deWit:1980gt}.
Using \eqref{charged hyper}, the gauge choice $A^i_{\;\alpha}\propto \delta^i_{\alpha}$ and 
substituting the expression for $\eta$ (\ref{gauge choice}) the SUSY equation becomes
\beq
\Big[2g(H\cdot P)
-2gi \gamma_5(J\cdot P)
-\frac{i}{L}\gamma_5
\Big]A^\alpha_i\sigma^i_{3j}\xi^j
-iF_i^{\;\;\alpha}\xi^i=0
\eeq
where we used the notation $J\cdot P= J^IP_I$. Since we want this equation to be valid for any Killing spinor $\xi$ we need
\beq
\label{cond SUSY hyper}
F_i^{\;\;\alpha}=-i2g {A^\alpha_j\sigma^j_{3i}}(H\cdot P)=0,
\qquad
2g (J\cdot P)=-\frac{1}{L}
\eeq

\subsection{An Attractor Solution}

In the following we show that the solution to the SUSY equations obtained in the last section are 
consistent with the equations of motion. We further show that all the fields 
become completely determined in terms of the {$AdS_{4}$ scale $L$} 
resulting in an attractor phenomenon for the scalar fields. The scale $L$ in turn is related to the flux $N$  by\footnote{We define the internal space metric as $ds^2_{internal}=L^2ds^2_{X_7}$ so that $\vol(X_7)$ does not carry powers of $L$.} \cite{Aharony:2008ug}
\beq
N=\frac{6L^6 \vol(X_7)}{(2\pi l_p)^6} \, .
\eeq

The  two derivative off-shell action for the bosonic fields is
\bal\label{Physical}
S&=\int d^4x\sqrt g\bigg[N_{IJ}\bar{X}^IX^J\left(\frac{R}{6}+D\right)+N_{IJ}\partial\bar{X}^I\partial X^J-\frac{1}{8}N_{IJ}Y^{ij I}Y^J_{ij}+{}\\
&\quad
\left(-\nabla A^{i}_{\;\beta}\nabla A_i^{\;\alpha}-\left(\frac{R}{6}-\frac{D}{2}\right) A^{i}_{\;\beta} A_i^{\;\alpha}+F^i_{\;\beta}F_i^{\;\alpha}+4g^2A^i_{\;\beta}\bar{X}^{\alpha}_{\;\gamma}X^{\gamma}_{\;\delta}A_i^{\;\delta}+gA^i_{\;\beta}Y^{jk\alpha}_{\;\;\gamma}A_k^{\;\gamma}\epsilon_{ij}\right)d_{\alpha}^{\;\,\beta}\bigg]
\eal
where $N_{IJ}:=(F_{IJ}-\bar{F_{IJ}})/2i$  with $F_{IJ} := \partial_I\partial_JF(X)$ and we have used the definition for the auxiliary field $F_i^{\;\;\alpha}= aA_i^{(z)\alpha}$ in the action (\ref{hypers}). Further details can be found in \cite{deWit:1984px,deWit:1980gt}.

The field $D$ acts as a Lagrange multiplier which then gives the equation of motion
\beq\label{D eq}
N_{IJ}\bar{X}^IX^J+\frac{1}{2}A^{i}_{\;\;\beta} A_i^{\;\;\alpha}d_{\alpha}^{\;\;\beta}=0
\eeq
At the same time we fix the coefficient of the Einstein-Hilbert term to have the canonical form
\beq\label{fix EH}
\frac{1}{6}N_{IJ}\bar{X}^IX^J-\frac{1}{6} A^{i}_{\;\;\beta} A_i^{\;\;\alpha}d_{\alpha}^{\;\;\beta}=\frac{1}{16\pi G}
\eeq
with $G$ the four dimensional Newton's constant. In the off-shell computation both equalities (\ref{D eq}) and (\ref{fix EH}) are valid only at infinity where the scalars approach their constant boundary value.

Solving both (\ref{D eq}) and (\ref{fix EH}) equations we get 
\beq\label{scalar constraint}
N_{IJ}\bar{X}^IX^J=\frac{1}{8\pi G},
\qquad
A^{i}_{\;\;\beta} A_i^{\;\;\alpha}d_{\alpha}^{\;\;\beta}=-\frac{1}{4\pi G}
\eeq
For the gauge $A_i^{\;\;\alpha}=\phi\delta_i^{\alpha}$ this gives 
\beq
\phi=\frac{1}{\sqrt{8\pi G}}
\eeq
after using $d_{\alpha}^{\;\;\alpha}=-2$. In Poincare supergravity the equations (\ref{scalar constraint}) are gauge fixing conditions. Here these are only valid at asymptotic infinity so we don't impose any further constraint on the scalars. 

The remaining equations of motion are computed to give
\begin{subequations}
\label{EOM}
\begin{align}
\mathbf{Y}:&\quad
N_{IJ}Y^{J1}{}_{1}=\frac{gi}{2\pi G}P_I, \label{eq1}
\qquad
Y^{I1}{}_{2}=Y^{I2}{}_{1}=0,\\
\mathbf{X}:&\quad
-\frac{2}{L^2}N_{IJ}\bar{X}^J-\frac{2}{L^2}\partial_I N_{HK}X^H\bar{X}^K
+\frac{1}{4}\partial_I N_{KH}Y^{K 1}{}_{1}Y^{H 1}{}_{ 1}+\frac{g^2}{\pi G}\bar{X}^K P_K P_I=0\label{eq 2}\\
\mathbf{A}^2:&\quad
-\frac{4}{L^2} +8g^2(J\cdot P)^2-\frac{4g}{L} (J\cdot P)=0\\
\mathbf{F}:&\quad
F_i^{\;\;\alpha}=0\\
\mathbf{G}_{\mu\nu}:&\quad
\frac{1}{4}R + 8g^2(J\cdot P)^2-\frac{4g}{L}(J\cdot P)-\frac{1}{L^2}=0,
\label{eq5}
\end{align}
\end{subequations}
In deriving these equations we have used the fact that $R(AdS_4)=-12/L^2$ and $Y^{1}_{\;\;1}=-Y^{2}_{\;\;2}=-2iJ/L$ at intermediate steps. Contracting equation (\ref{eq 2}) with $X^I$ we deduce
\beq
N_{IJ}\bar{X}^IX^J=\frac{L^2g^2}{2\pi G}(\bar{X}^IP_I)(X^IP_I)
\eeq
where we used the fact that $X^I\partial_IN_{HK}X^H\bar{X}^K=0$, valid for any homogeneous function $F(X)$ with weyl weight 2. Together with equation (\ref{eq1}) we have 
\beq
4g^2(J\cdot P)^2+\frac{2g}{L}(J\cdot P)=0,\qquad
8g^2(J\cdot P)^2=\frac{2}{L^2}.
\eeq
This gives back the condition (\ref{cond SUSY hyper})
\beq
2g(J\cdot P)=-\frac{1}{L}
\eeq
which is also consistent with Einstein's equation (\ref{eq5}).
Note that equation (\ref{eq 2}) imposes additional constraints on the scalars. 

We now apply this general formalism to our model with the prepotential \eqref{Ourprepot}. 
We start by fixing the values of the scalars at asymptotic infinity.  
Equation (\ref{scalar constraint})  becomes
\beq
\frac{1}{4 i}|X^0|^2\left(\sqrt{\frac{X^1}{X^0}}-\bar{\sqrt{\frac{X^1}{X^0}}}\right)^3
=\frac{1}{8\pi G}
\qquad
\Leftrightarrow\qquad
(J^0)^{1/2}(J^1)^{3/2}=\frac{i}{16\pi G} 
\label{NijXiXj}.
\eeq
where we have chosen the branch cut of $\sqrt{X^1/X^0}$ along the negative real line and $\sqrt{-1}=-i$.\footnote{This choice is justified by the fact that we are taking $J^0>0$ and $J^1<0$.}

Equations (\ref{eq1}) and (\ref{eq 2}) give the attractor equations
\beq\label{attract values}
8gLJ^0P_0=-1,
\qquad
8gLJ^1P_1=-3
\eeq
which will be used later on to show that the renormalized action depends only on the size of $AdS_4$ through the scaling variable $z$ in \eqref{zdef} related to the radius of curvature of $AdS_{4}$ in the units of 4d Planck length \eqref{F-ABJM}. 

\section{Localization in Bulk Supergravity  and the Airy Function\label{sec:Localization}}

In order to use localization we need a ``real'' fermionic symmetry that squares to a compact $U(1)$ together with possible gauge transformations, that is,
\beq
\delta^2=\mathcal{L}_{U(1)}+G.
\eeq
For the problem at hand we choose  the Killing spinor (\ref{Killing sp})
\beq\label{Killing sp localization}
\xi=\left(\begin{array}{c}
\chi_+\times \epsilon^1_-\\
	(\sigma_3\chi_+)\times \epsilon^2_-
	  \end{array}
\right),
\eeq
normalized so that $\xi^{\dagger}\xi=\cosh(\eta)$. The associated Killing vector $v$
\beq
v=\xi^{\dagger}\gamma^{\mu}\xi\partial_{\mu}=2\epsilon_-^{\dagger}\gamma^i\epsilon_-\partial_i
\eeq
generates translations along the Hopf fiber of $S^3$. Using the supersymmetric transformations of the vector fields we can check that the fermionic symmetry generated by that Killing spinor is in fact a coordinate translation plus a gauge transformation. This is already guaranteed by the off-shell closure of the superconformal algebra. However, note that our choice for $\eta$ (\ref{gauge choice}) is consistent with the localization principle in the sense that it doesn't generate conformal transformations since ${\xi}^\dagger \eta=0$. Therefore a deformation to the path integral of the form 
\beq 
\delta S=-t\delta((\delta\psi)^{\dagger}\psi)
\eeq
with $\psi$ any fermion field in the theory, is exact. The bosonic part of this deformation acts as a regulator and dominates the path integral in the limit $t\rightarrow \infty$. In this limit the theory projects onto the saddle points of the deformation and the semiclassical aproximation becomes exact. This is the localization principle. Since the bosonic action is a positive definite quantity, the saddles are determined by the BPS equations
\beq \label{BPS loc eq}
\delta_{\xi}\psi=0.
\eeq
These are coupled first order differential equations we need to solve on the $AdS_4$ background with certain boundary conditions. In appendix~\ref{Loc appendix} we construct the bosonic part of the localization action. 
We find the instanton solution
\bea\label{Loc solutions}
X^{I}&=& H^{I}+iJ^{I}=iJ^{I}+\frac{J^{I } h^{I}}{\cosh(\eta)}\, , \\
(Y^{I})^{1}_{\;\;1}&=& -(Y^{I})^{2}_{\;\;2}=-2i\frac{J^{I}}{L}+2\frac{J^{I}h^{I}}{L\cosh^2(\eta)} \, ,  \qquad (I=0, 1)
\eea
where $J^{I}$ are fixed to the attractor values \eqn{NijXiXj}, \eqn{attract values} and $h^{I}$ are real numbers that parametrize the solutions, and there is no summation over the index $I$ on the right hand side of these equations. Note that the auxiliary fields $Y^{I}$ have a non-trivial profile which allows for the scalar fields $X^{I}$ to climb above the attractor value. This is a common feature of localization in supersymmetric field theories. The  space of solutions is therefore $\mathbb{R}^{n_v+1}$, with $n_v$ the number of vector multiplets and in our case $n_{v}=1$. These solutions are also reminiscent of the localization problem on $AdS_2\times S^2$ relevant for black hole entropy \cite{Dabholkar:2010uh,Gomes:2013cca,Gupta:2012cy}. 

In the hypermultiplet sector the story is a bit more complicated,  because the off-shell extension of the hypermultiplet requires considering an infinite set of hypermultiplets as explained before in~\S\ref{Hypermultiplet sec}. An important obstacle to localization of the off-shell hypermultiplet is that it requires solving the BPS equation (\ref{BPS loc eq}) under some constraints. To circumvent this problem we proceed in a different way. We solve the off-shell hyper susy equations for all Killing spinors and interpret the off-shell solution as a background for the hypers rather then an off-shell saddle point of an exact deformation. We believe that both point of views should be equivalent but right now we lack a clear technical understanding. We find (check appendix~\ref{Loc appendix})
\beq
F_i^{\alpha}=\frac{2g}{\sqrt{8\pi G}}H^IP_I\sigma^{\alpha}_{3i},\qquad
A_i^{\;\;\alpha}=\frac{1}{\sqrt{8\pi G}}\delta^{\alpha}_i
\eeq with $H$ given as in (\ref{Loc solutions}).

In this gauge only the scalar fields in the vector multiplets are excited for the off-shell localizing solution and the metric is held fixed.  However, the Weyl-invariant physical metric depends on the conformal factor constructed from the scalar fields which is precisely $N_{IJ}X^{I} \bar{X}^{J}$ and this has a nontrivial profile for the localizing solution.

\subsection{Action on the Localization Locus}
\label{sec:action}

In this subsection we turn to the evaluation of  the physical action \eqref{Physical} on these localizing  solutions.  The resulting expression is quite complicated but after manipulations in mathematica can be seen to be integrable. Some of the intermediate steps are described in the appendix~\ref{app:Action}. Below we present the main results.

Consider  first the action for the gravity multiplet coupled to vector multiplets
evaluated  on the localization saddles (\ref{Loc solutions}):
\beq\label{vectoraction}
S_\text{vec}=\int d^4x\sqrt g\bigg[N_{IJ}\bar{X}^IX^J\frac{R}{6}+N_{IJ}\partial\bar{X}^I\partial X^J-\frac{1}{8}N_{IJ}Y^{ij I}Y^J_{ij}\bigg]_{\text{Loc. locus}} \, .
\eeq
It is convenient to define the coordinate $r =\cosh(\eta)$ so that  $r=1$ is the center of $AdS_{4}$  and $r \rightarrow \infty$ is the boundary. At very large $r$, the integral diverges and we regularize it with a cutoff $r_0$. 
Thus the $r$ integral has two boundaries, one at   $r=1$  and the other at $r=r_{0}$. 
Up  to terms that vanish faster than $\cO(1/r_0)$, we obtain
\bal
\label{reg-vec}
S_\text{vec}=-\frac{\Omega_3L^2}{32\pi G}\Big[&4r_0^3
+\frac{r_0}{2}(3 (h^{1})^2+6 h^{1}h^0-(h^0)^2-24)+2i r_0(3h^{1}+h^0)
\\&{}
+8(1-ih^{1})^{3/2}\sqrt{1-ih^0}\Big]
\eal
where we have used $\sqrt{J^0 (J^{1})^3}/2i=1/32\pi G$ (\ref{NijXiXj}).

Now consider the action for the hypermultiplet sector \eqn{hypers}:
\beq\label{hyperaction}
S_\text{hyp}=\int d^{4}x{\sqrt{g}}\left[ -\frac{1}{6}R A^2+\left(4g^2A^i_{\;\;\beta}\bar{X}^{\alpha}_{\;\;\gamma}X^{\gamma}_{\;\;\delta}A_i^{\;\;\delta}+gA^i_{\;\;\beta}Y^{jk\alpha}_{\;\;\;\gamma}A_k^{\;\;\gamma}\epsilon_{ij}+F_i^{\;\;\alpha}F^i_{\;\;\beta}\right)d_{\alpha}^{\;\;\beta}\right]
\eeq
Note that this action vanishes at the on-shell level because being proportional to $A^2$, the proportionality factor, at on-shell level, equals the equations of motion. For our off-shell localizing solutions, on the other hand, we obtain
\beq
\label{hyp-action-value}
S_\text{hyp}=i\frac{\Omega_3L^2}{16\pi G}(r_0-2)(h^0+3h^{1})+\mathcal{O}(1/r_0)
\eeq
The divergence in this term is cancelled precisely by the term linear in $h$ and $h^0$ coming from \eqn{reg-vec}  
which is what one expects if the   variational problem is well-defined. We are thus left with the total action
\beq
\label{reg-total}
S=-\frac{\Omega_3L^2}{32\pi G}\left[4r_0^3
+\frac{r_0}{2}\left(3 (h^{1})^2+6 h^{1 }h^0-(h^0)^2-24\right)
+8(1-ih^{1})^{3/2}\sqrt{1-ih^0}
+4i(3h^{1}+h^0)\right].
\eeq

\subsection{Holographic Renormalization and Flux Boundary Term}

In this section we construct the boundary counterterms that remove cuttoff $r_0$ dependent terms in the action \eqn{reg-total}. This is the usual procedure of holographic renormalization \cite{Bianchi:2001kw,Emparan:1999pm}. We show that these terms do not contribute with additional $h^0,h$ dependent terms to the renormalized action. This is a priori possible since the scalars decay at infinity as $1/r_0$ with $r_0$ the IR cuttoff and therefore a suitable local combination of these with other boundary scalar quantities could result in $h^0,h$ dependence. These counterterms should arise naturally from supersymmetric considerations. In this work nevertheless we only consider bosonic counterterms without caring about their supersymmetric completion.
 
In terms of the cuttoff $r_0$ the divergence goes as
\beq \label{divergence}
\frac{\Omega_3 L^2}{8\pi G}\left[
-r_0^3+r_0\left(3-\frac{3( h^{1})^2+6 h^{1}h^0-(h^0)^2}{8}\right)\right]
\eeq
To remove the IR divergence (\ref{divergence}) we can add the boundary action 
\beq
S_{\text{counter-term}}=S_\text{GH}+\frac{1}{2}S_{B2}
\eeq
where $S_\text{GH}$ is the Gibbons-Hawking term obtained from varying the Ricci scalar in the 
vector multiplet action
\bal\label{bnd GH}
S_\text{GH}
&=\int d^3x\,\sqrt{g_3}
N_{IJ}\bar{X}^IX^J\frac{\kappa}{3}
=\Omega_3L^2\frac{\sqrt{J^0(J)^3}}{2i}\left(1+\frac{3(h^{1})^2+6h^{1}h^0-(h^0)^2}{8r_0^2}\right)r_0(r_0^2-1)
\\&=
\frac{\Omega_3 L^2}{8\pi G}\left[r_0^3-r_0\left(1-\frac{3(h^{1})^2+6h^{1}h^0-(h^0)^2}{8}\right)\right],
\eal
with the trace of the extrinsic curvature given by {$\kappa=3\coth(\eta_0)/L$}, 
and $S_{B2}$ is the boundary term that is proportional to the scalar curvature of the boundary manifold 
$\cR=24/(L^2\sinh^2(\eta))$
\bal
S_{B2}
&=-\int d^3x\,\sqrt{g_3}
N_{IJ}\bar{X}^IX^J\frac{L\cR}{6}
=-\Omega_3L^2\frac{\sqrt{J^0(J)^3}}{2i}\left(1+\frac{3(h^{1})^2+6h^{1}h^0-(h^0)^2}{8r_0^2}\right)4\sqrt{r_0^2-1}
\\&\sim
\frac{\Omega_3 L^2}{8\pi G}\left[-4r_0\right].
\eal
These two terms exactly cancel the divergence in \eqn{divergence}. 
However, the equivalent term stemming from the curvature coupling in the 
hypermultiplet action reintroduces the divergence
\beq
\label{remaining div}
\int d^3x\,\sqrt{g_3}
(-A^2)\left(\frac{\kappa}{3}-\frac{\cR}{12}\right)
=\Omega_3L^2\frac{-1}{4\pi G}
\left(r_0(r_0^2-1)-2\sqrt{r_0^2-1}\right)
\sim-\frac{\Omega_3L^2}{4\pi G}(r_0^3-3r_0)\,.
\eeq

Besides these boundary terms we should consider a new kind of boundary term, one of topological nature. 
In M-theory language this boundary term can be seen to source $N$ units of electric flux  for the 3-form field $C_3$ 
that couples to the M2-branes forcing us to work in the microcanonical ensemble. This discussion is very similar to what 
happens in the $AdS_2$ case, where the electric field is non-normalizable and has to be fixed in the path integral 
while the chemical potential can fluctuate  \cite{Sen:2008vm}. To guarantee that the equations of motion are 
satisfied also at the boundary one inserts boundary Wilson lines. 
In the $AdS_4$ case, the $C_3$ field is given in the coordinates (\ref{AdS4 metric}) by
\beq
\label{C3}
F_4=a\,\omega_{AdS_4}
\quad\Rightarrow\quad
C_3=\left[a\left(\frac{1}{3}\cosh^3(\eta)-\cosh(\eta)\right)+b\right]d\Omega_3
\eeq
where $a,b$ are constants. 
We thus see that the $C_3$ field suffers from the same behaviour as in the $AdS_2$ case - 
its flux is non-normalizable at the boundary of $AdS_4$ and has to be fixed in the path integral. 
By the same token we should allow the mode $b$ to fluctuate. However under the M-theory truncation 
to four dimensions the flux becomes non-dynamical, as we use its equations of motion, and we do not 
see it in the formalism developed before.

To determine the correct boundary term for the $C_3$ flux we consider the truncated on-shell action in four 
dimensions as a function of its field strength $F_4$. A naive reduction from eleven to four 
dimensions would result in an action
\beq \label{truncated}
\frac{L^7\vol(X_7)}{64\pi G_{11}}\int d^4x\sqrt{g_4}\left(R(g_4)+\frac{1}{4L^2}R(X_7)-\frac{1}{8\cdot4!}F_4^2\right)
\eeq
However plugging in the on-shell value of $F_4^2/4!=-36/L^2$ does not give the correct cosmological 
constant. It is well known that it is the action (\ref{truncated}) with the sign in $F^2$ flipped that gives 
a good truncation of the theory to four dimensions \cite{Gauntlett:2009zw}. The on-shell truncated action as a function of 
the field strength $F_4=dC_3$ is therefore
\beq
\frac{1}{16\pi G_{11}}\int_{AdS_4\times X_7} \frac{2}{3}\frac{F_{4}^2}{4!}
\eeq
On the other hand, the on-shell value of $C_3$ is given by \eqn{C3} with $a=3iL^3/8$, such that the 
flux integrates to $N$
\beq\label{F4 flux}
\int_{X_7} \star F_4=i6L^6\vol(X_7)=i(2\pi)^6N.
\eeq
Regularity of $C_3$ at $\eta=0$, where the size of $S^3$ shrinks to zero, determines the value of $b$ to be $b=2a/3$.

As stated above, from a holographic point of view we want to keep the non-normalizable mode $a$ fixed 
and integrate over the normalizable one, $b$. 
A correct variational principle\footnote{Starting from $\int F_4\wedge \star F_4$ a variation of $C_3$ gives 
$\sim \int d(C_3\wedge \star F_4)$ from which follows the boundary term.} 
for the $C_3$ field therefore requires adding the boundary term
\beq\label{flux bnd term}
-i\frac{N}{3\pi^2}\int_{S^3} C_3
\eeq
The divergence of this topological term exactly cancels the remaining divergence 
from the boundary contribution to the hypermultiplet action \eqn{remaining div}.
The remaining finite part of \eqn{flux bnd term} then modifies the renormalized action (\ref{reg-total}), that is,
\beq\label{additional cntrb renorm}
S
\quad\rightarrow\quad
S+\frac{\pi 2\sqrt{2}}{3}k^{1/2}N^{3/2}
\eeq
where we have used that
\beq\label{G4}
\frac{L^7 \vol(X_7)}{4\cdot16\pi G_{11}}=\frac{L^7 \vol(X_7)}{4\cdot(2\pi)^8}=\frac{1}{16\pi G_4}
\eeq
in units where $l_P=1$ and $\vol(X_7)\sim \vol(M_6)/k$.

\subsection{The Final Integral and the Airy Function}

After regularization the renormalized action becomes
\beq\label{Ren S}
S_{ren}=-\frac{\pi \sqrt{2}}{3}k^{1/2}N^{3/2}\left[(1-ih^{1})^{3/2}\sqrt{1-ih^0}
+\frac{i}{2}(3h^{1}+h^0)-2\right]
\eeq 
where the constant contribution in the formula above comes from the flux boundary term (\ref{additional cntrb renorm}). We have used the formula
\beq
\frac{\Omega_3 L^2}{4\pi G}=\frac{\pi \sqrt{2}}{3}k^{1/2}N^{3/2}
\eeq with the help of (\ref{F4 flux}) and (\ref{G4}).

It is convenient to define new variables $\phi^{I}$ as
\beq
\phi^{0} := \frac{\pi}{3\sqrt{2}}\frac{N^{3/2}}{k^{1/2}}(1 -i h^{0}),
\qquad
\phi^{1}:=\frac{\pi}{\sqrt{2}}k^{1/2}N^{1/2}(1-ih^{1}).
\eeq
Up to a proportionality factor these are the values of the fields $X^{I}$ for the localizing solutions at the center of $AdS_{4}$ \eqn{Loc solutions}. The renormalized action (\ref{Ren S}) becomes simply
\beq\label{ren-action}
S_{ren}= -\frac{2\sqrt{2}}{\pi\sqrt{3}} \sqrt{\phi^0(\phi^1)^3}+N\phi^{1} +k\phi^{0} \, .
\eeq
We thus see that  the localization integral looks precisely like a Laplace transform of the partition function
\beq
Z(\phi) = e^{F(\phi)},
\qquad\text{with}\quad
F(\phi)=\frac{2\sqrt{2}}{\pi\sqrt{3}} \sqrt{\phi^0(\phi^1)^3}.
\eeq
Note that the euclidean path integral is weighted with $e^{-S_{ren}}$. With these variables it also makes clear that we are working in a microcanonical ensemble with $N,k$ the two ``electric charges'' of the problem. From the eleven dimensional point of view $N$ being the flux on $AdS_4$ is effectively ellectric while the charge $k$ is magnetic because it is the flux of the Hopf fiber. This apparent contradiction is explained by noticing that the theory where the prepotential is the ``square root'' one is obtained from the original truncation after a symplectic transformation. This being an electric-magnetic duality transformation turns $k$ to be electric.
After massaging equation (\ref{ren-action}) we obtain
\beq\label{Sren gaussian}
-S_{ren}=-k\left(\sqrt{\phi^0}-\frac{1}{\pi k}\sqrt{\frac{2}{3}}(\phi^1)^{3/2}\right)^2
+\frac{2}{3\pi^2 k}(\phi^1)^3-N\phi^1
\eeq
Note that $2(\phi^1)^3/3\pi^2k$ is just the grand canonical potential (\ref{grand canonical pot}) in variables where $ \phi^1=\mu$. 

Since the renormalized action appears in the exponent of a functional integral, the leading large $N$ behavior can be extracted easily in the saddle point approximation. For this purpose it is not necessary to know the full measure on the collective coordinate manifold. It is sufficient to evaluate the renormalized action at the extremum of the function of two variables $\phi^0$ and $\phi^1$ in (\ref{Sren gaussian}). The resulting expression for the free energy matches  with the expression that is usually obtained on the gravity side by evaluating the \textit{on shell} action.
Note that our computation is around a very different \textit{off-shell} configurations that has nothing to do with the classical equations of motion of the on-shell theory and the computation of the renormalized action involves a complicated integral over space-time coordinates. It is thus a useful check   that one obtains a simple analytic expression for the renormalized action with the correct behavior at the saddle point.

If we can integrate explicitly the gaussian in (\ref{Sren gaussian}), for which we assume  a flat measure for the variables $(u=\sqrt{\phi^0},\mu=\phi^1)$, then the final expression for the integral is
\beq
Z_{ABJM}=\int e^{-S}=\int_{\gamma-i\infty}^{\gamma+i\infty} d\mu \exp\left({\frac{2}{3\pi^2k}\mu^3- N\mu}\right)
\eeq
where $\gamma=\pi k^{1/2}N^{1/2}/\sqrt{2}$, which we identify with the gauge theory computation (\ref{airy}) in section~\ref{sec:MM}.
After deforming the contour\footnote{We start by deforming the contour $\{1-i\infty,1+i\infty\}$ to $\{\epsilon-i\infty,\epsilon+i\infty\}$, with $0<\epsilon\ \ll 1$. Then it becomes easy to show that we can deform the contour to a triangle shape within the angles $]\frac{\pi}{2},\frac{\pi}{6}[$ in the upper half and $]-\frac{\pi}{2},-\frac{\pi}{6}[$ in the lower half.} we can rewrite, up to a normalization factor, $Z_{ABJM}$ as 
\begin{equation}\label{Airy fnct}
Z_{ABJM}=\int_{\infty e^{-i\frac{\pi}{3}}}^{\infty e^{+i\frac{\pi}{3}} } dt \,\exp\left[\frac{1}{3}t^3-zt\right]=\Ai(z),\;\;z=(\pi^2 k/2)^{1/3}N=z_{\text{ABJM}}
\end{equation} which is the exact answer up to $1/N$ corrections in $z_{ABJM}$. 

Our results are reminiscent  of a similar story for BPS black holes and $AdS_{2}$ holography. In that context, it was conjectured  \cite{Ooguri:2004zv} that the exact black hole degeneracies are related to an appropriate Laplace transform of $|Z_{top}|^{2}$ where $Z_{top}$ is the partition function of the topological string associated with the Calabi-Yau manifold of compactification. Using localization techniques for the $AdS_{2} \times S^{2}$ near horizon background, it was shown in \cite{Dabholkar:2010uh}, that the integrand for localization integral is indeed proportional to $|Z_{top}|^{2}$ if one ignores nonperturbative corrections. In the present context of $AdS_{4}$, we are obtaining an integrand that is proportional to the \textit{holomorphic} $Z_{top}$ but with just the tree level square-root prepotential. A connection to topological string in this case  is not immediately clear because  we have a gauged supergravity obtained by a consistent truncation rather than an ungauged supergravity obtained by  a compactification on a Calabi-Yau manifold. However, the following observations suggest a tantalizing connection.

To be concrete, we consider the best-understood case of the ABJM theory which corresponds to a truncation of Type-IIA on $\mathbb{CP}^{3}$.
In the boundary gauge theory, the Airy function has a natural representation as a Laplace-like symplectic transform of the  tree-level cubic prepotential of   the topological string on local $\mathbb{CP}^{1} \times \mathbb{CP}^{1}$ in the large-radius frame \cite{mp-fermi}.   It is striking that our computation from localization in the \textit{bulk} supergravity yields a very similar integral representation for the Airy function, and moreover the square-root potential is related to the cubic one by a symplectic transformation. It is natural  to ask which quantity  provides the nonperturbative completion  of the tree-level prepotential. Such a prepotential  would then be the input for evaluating the renormalized action on our localizing solutions.  
 There is no \textit{a priori} notion of a topological string for $\mathbb{CP}^{3}$; however, the prepotential that determines the chiral effective action of the gauged supergravity is a well-defined object.  Now, it has been proposed, using the matrix model representation,  that the exact prepotential for the local $\mathbb{CP}^{1} \times \mathbb{CP}^{1}$  is computed by the refined topological string \cite{dmp1, dmp2, Hatsuda:2013oxa, Kallen:2013qla}. This answer would include corrections both from world-sheet instantons and Euclidean D2-brane instantons and would give the fully quantum corrected partition function for the ABJM theory which nonperturbatively completes the Airy function.
It is thus tempting to conjecture that the instanton-corrected exact prepotential of the refined topological string on local $\mathbb{CP}^{1} \times \mathbb{CP}^{1}$ has a physical interpretation as the   prepotential of the gauged supergravity obtained by a truncation on $\mathbb{CP}^{3}$.

\acknowledgments

We would like to acknowledge the hospitality of  the CERN theory group  where this project was initiated. N.D would  like to thank also the Weizmann Institute for hospitality. We are grateful to Davide Cassani, Nick Halmagyi, Amir Kashani-Poor, Albrecht Klemm, Marcos Mari\~no, Dario Martelli, Sameer Murthy, Sergio Ferrara and Stefan Vandoren for  discussions. We thank  Filippo Passerini for early collaboration.
The work of A.~D.~was conducted within the framework of  the ILP LABEX (ANR-10-LABX-63)  supported by French state funds managed by the ANR within the Investissements d'Avenir programme under reference ANR-11-IDEX-0004-02,  and by the project QHNS in the program ANR Blanc SIMI5 of the Agence National de la Recherche. The research of J.~G.~ has received funding from the European Research Council under the European Community's Seventh Framework Programme (FP7/2007-2013)/ERC grant agreement no. [247252].

\appendix 

\section{Killing spinors of the vacuum $AdS_{4}$ solution \label{Killing}}

In this appendix we describe the solutions to the  Killing spinor equations \eqref{Killing eq2}.  We use greek indices $\mu,\nu\ldots$ for the four dimensional coordinates and roman $a,b,i,j,\ldots$ for tangent space indices. We set $L=1$ in what follows.

For the metric (\ref{AdS4 metric}) we choose the vielbein
\beq
e^0=d\eta,
\qquad
e^i=\sinh(\eta)\bar{e}^i,
\qquad
\text{with}\quad\sum_i\bar{e}^i\bar{e}^i=d\Omega_3^2
\eeq
which give the spin connections%
\footnote{We have used  $de^{a}+\omega^{ab}\wedge e^b=0$.}
\beq
\omega^{i0}=\cosh(\eta)\bar{e}^i,\qquad
\omega^{ij}=\bar{\omega}^{ij}
\eeq

We choose the following gamma matrices representation
\beq
\gamma_0=\sigma_1\times \mathbb{I},
\qquad
\gamma_i=\sigma_2\times \sigma_i,
\qquad
\gamma_5=-\gamma_0\gamma_1\gamma_2\gamma_3=\sigma_3\times \mathbb{I}
\eeq
so that $\{\gamma_{a},\gamma_{b}\}=2\delta_{ab}$.

The Killing spinor equation is solved to give \cite{Lu:1998nu}
\beq\label{Killing spinors}
\xi^1=A\,\chi_+\times \epsilon_-+B\,\chi_-\times \epsilon_+\,,\qquad
\xi^2=C\,(\sigma_3\chi_+)\times \epsilon_-+D\,(\sigma_3\chi_-)\times \epsilon_+\
\eeq
with
\beq
\chi_+=\left(\begin{array}{c}
\sinh(\eta/2)\\
-i\cosh(\eta/2)
\end{array}\right),
\qquad 
\chi_-=\left(\begin{array}{c}
\cosh(\eta/2)\\
-i\sinh(\eta/2)
\end{array}\right)
\eeq
and $A,B,C,D$ complex constants. The $\epsilon_{\pm}$ are Killing spinors on the $S^3$, that is,
\beq
\nabla_{\hat{a}} \epsilon_{\pm}=\pm\frac{i}{2}\tilde{\gamma}_{\hat{a}}\epsilon_{\pm}\,,
\eeq
and $\tilde{\gamma}_{\hat{a}}$ are gamma matrices on $S^3$.
Since $S^3$ admits two Killing spinors we have for each $i=1,2$ component four Killing spinors. This gives a total of sixteen solutions that get reduced to eight after imposing a Majorana-symplectic reality constraint \cite{Cortes:2003zd}
\beq\label{reality cond}
(\xi^i)^{*}=-i\epsilon_{ij}(\mathbb{I}\times \sigma_2)\xi^j.
\eeq
In the following we construct explicitly this basis. First choose Killing spinors $\epsilon^i_{\pm}$ in $S^3$ space so that \cite{Lu:1998nu}
\beq
(\epsilon^i_{\pm})^*=-i\epsilon_{ij}\sigma_2 \epsilon^j_{\pm}.
\eeq
Under the reality constraint (\ref{reality cond}) we construct the ``real'' combinations
\beq\label{Killing sp}
\left(\begin{array}{c}
\xi^1\\
	\xi^2
\end{array}
\right)=\left(\begin{array}{c}
\chi_+\times \epsilon^1_-\\
	(\sigma_3\chi_+)\times \epsilon^2_-
\end{array}
\right)
,\;\left(\begin{array}{c}
\chi_-\times \epsilon^1_+\\
	(\sigma_3\chi_-)\times \epsilon^2_+
\end{array}
\right),\;\left(\begin{array}{c}
i\chi_+\times \epsilon^2_-\\
	i(\sigma_3\chi_+)\times \epsilon^1_-
\end{array}
\right),\;\left(\begin{array}{c}
i\chi_-\times \epsilon^2_+\\
	i(\sigma_3\chi_-)\times \epsilon^1_+
\end{array}
\right)
\eeq
together with the imaginary ones
\beq
\left(\begin{array}{c}
\xi^1\\
	\xi^2
\end{array}
\right)=\left(\begin{array}{c}
i\chi_+\times \epsilon^1_-\\
	-i(\sigma_3\chi_+)\times \epsilon^2_-
\end{array}
\right)
,\;\left(\begin{array}{c}
i\chi_-\times \epsilon^1_+\\
	-i(\sigma_3\chi_-)\times \epsilon^2_+
\end{array}
\right),\;\left(\begin{array}{c}
\chi_+\times \epsilon^2_-\\
	-(\sigma_3\chi_+)\times \epsilon^1_-
\end{array}
\right),\;\left(\begin{array}{c}
\chi_-\times \epsilon^2_+\\
	-(\sigma_3\chi_-)\times \epsilon^1_+
\end{array}
\right)
\eeq
which give a total of eight Killing spinors.

\section{Localization action}\label{Loc appendix}

In this section we compute the bosonic part of the localization action and solve the localizations equations that are obtained from it.

Firstly we describe some properties of the Killing spinor we will use for localization. We have choosen (\ref{Killing sp localization}) 
\beq
\xi=\left(\begin{array}{c}
          \chi_+\times \epsilon^1_-\\
	(\sigma_3\chi_+)\times \epsilon^2_-
	     \end{array}
\right),
\eeq where 
\beq
\chi_+=\left(\begin{array}{c}
      \sinh(\eta/2)\\
	-i\cosh(\eta/2)
     \end{array}\right),
\eeq and $\epsilon_-$ obeys a symplectic-majorana reality condition
\beq
(\epsilon^i_{-})^*=-i\epsilon_{ij}\sigma_2 \epsilon^j_{-}.
\eeq We normalize the Killing spinor so that
\begin{eqnarray}
&& \xi^{\dagger}\xi=\sum_i (\xi^{i})^{\dagger}\xi^i=\cosh(\eta)\\
&& \xi^{\dagger}\gamma^{0}\xi=0\,,
\quad
\xi^{\dagger}\gamma^{i}\xi=\sum_i (\xi^{i})^{\dagger}\gamma^{i}\xi^i=V^{i},
\quad
i\in\, S^3,
\qquad
V^2=\sinh(\eta)^2
\end{eqnarray}where $V^{i}$ is a right-invariant\footnote{The isometry of $S^3$ is $SU(2)_L\times SU(2)_R$ from which we can define left/right invariant Killing vectors.} Killing vector. In the following we present some properties of the Killing spinor that will be useful for computing the localization action
\begin{eqnarray}
&& (\xi^1)^{\dagger}\xi^2=0\\
&&\xi^{\dagger}\gamma_5\xi=-1\\
&&\xi^{\dagger}\gamma^{ab}\xi=0,\quad a,b=0\ldots 5 \\
&& \xi^{\dagger}\eta=\xi^{\dagger}\gamma_{\mu}\gamma_5\eta=0\\
&&\xi^{\dagger}\gamma_{\mu}\eta=-i\sinh(\eta)\delta_{\mu,0}
\end{eqnarray}The bosonic part of the localization action is given by the square of the fermionic SUSY transformation $(\delta\Omega)^{\dagger}\delta\Omega$ with
\beq
\delta \Omega^{i}=\frac{1}{2}F_{\mu\nu}\gamma^{\mu\nu}\xi^i + 2i\slashed D(H-i\gamma_5 J)\xi^{i}+Y^i_{\;\;j}\xi^j-2(H+i\gamma_5 J)\eta^i
\eeq and we have parametrized the scalars as $X=H+iJ$. We offset the values of the fields so that they approach zero at the boundary. Due to the symmetry of the problem we take $Y^1_{\;\;2}=Y^2_{\;\;1}=0$.

After some tedious algebra the bosonic part of the localization action can be written as a sum of perfect squares
\begin{eqnarray}
(\delta \Omega)^{\dagger}\delta \Omega&=& \frac{1}{4\cosh(\eta)}\left(F_{ab}\cosh(\eta)-\frac{1}{2}\epsilon_{abcd}F^{cd}-2\epsilon_{abcd}\partial^{c}JV^d-2\Theta_{ab}J\right)^2+\nonumber\\
&&{}+\frac{1}{2\cosh(\eta)}\left(-\frac{1}{2}\epsilon_{abcd}F^{bc}V^d+2\partial_{a}(J\cosh(\eta))\right)^2
+ \frac{2}{\cosh(\eta)}\left(V^a\partial_aJ\right)^2
\nonumber \\
&&{}+\frac{1}{2\cosh(\eta)}\left(F_{ab}V^b-2\partial_aJ\right)^2
+4\cosh(\eta)\left(\partial_iH\right)^2+4\cosh(\eta)\left(\partial_{\eta}H+\frac{\sinh(\eta)}{\cosh(\eta)}H\right)^2\nonumber \\
&&{}+ \cosh(\eta)\left(Y^1_{\;\;1}-\frac{2H}{\cosh(\eta)}\right)^2
\end{eqnarray}where we defined the tensor $\Theta_{ab}$ as
\beq
\Theta_{ab}=-i\xi^{\dagger}\gamma_{ab}\gamma_5\eta,
\eeq which is real and has the following properties which are useful in computing the localization action
\beq
\Theta_{0i}=-\frac{V_i}{\sinh(\eta)},
\qquad
\Theta_{ij}=\frac{\cosh(\eta)}{\sinh(\eta)}\epsilon_{ijk}V^k,\quad i,j\in S^3.
\eeq
Since the localization action must vanish on the saddle points, this locus is alternatively determined by the vanishing of the different squares. 

The equations for $H$ and $Y$ can be easily solved to give
\beq
H=\frac{C}{\cosh(\eta)},
\qquad
Y^1_{\;\;1}=\frac{2C}{\cosh(\eta)^2}
\eeq
with $C$ an arbitrary constant. If we had used instead the real Killing spinor with $\chi_{-}$ we would have found the same solution for $X$ but now with $Y^1_{\;\;1}=-\frac{2C}{\cosh(\eta)^2}$. This would have implied that in order to obey all the Killing spinor equations we would need $C=0$ and therefore $X$ had to be constant. Note that the localization action remains the same after flipping the sign of $h^0,h^1$ in the renormalized action (\ref{Ren S}).

In order to solve for $J$ and $F_{ab}$ it requires a bit more work. Lets first look at the equations
\begin{eqnarray}
V^{\mu}\partial_{\mu}J=0,
\qquad
F_{\mu\nu}V^{\nu}=2\partial_{\mu}J.
\end{eqnarray}In a gauge where $V^{\mu}A_{\mu}=0$, which we can take assuming there is no Wilson line along that direction, we can show that the above equations are equivalent to 
\begin{equation}
\partial_zJ=0,
\qquad
\partial_zA_{\mu}=2\partial_{\mu}J
\end{equation}in coordinates where $V^{\mu}\partial_{\mu}=\partial_z$ \footnote{Since $V^{\mu}$ is a Killing vector we can always find coordinates where the metric looks like $ds^2=(dz+A)^2+g_{\mu\nu}dx^{\mu\nu}$ with both $A$ and $g_{\mu\nu}$ independent of $z$. In this particular problem this vector generates translations on the Hopf fiber of the $S^3$.}. These can be solved to give 
\begin{equation}
\partial_z^2A_{\mu}=0
\quad\Rightarrow\quad A_{\mu}=az+b
\end{equation}where $a,b$ are functions independent of $z$. Since $z$ is parametrizes a circle we must have $a=0$ otherwise the gauge field won't be periodic. Therefore we conclude that
\begin{equation}
\partial_{\mu}J=0
\quad\Rightarrow\quad
J=0,
\quad
F_{\mu\nu}V^{\nu}=0
\end{equation} 

To determine completely the gauge field we need to solve the equation
\beq
F_{ab}\cosh(\eta)-\frac{1}{2}\epsilon_{abcd}F^{cd}=0
\quad\Leftrightarrow\quad
\cosh(\eta)F=\star F
\eeq This gives immediately $F=0$. Note that at the origin $\eta=0$, this equation admits a self-dual instanton solution.

In order to make the superconformal algebra close off-shell in the hypermultiplet sector we need to consider an infinite family of hypers $(A_i^{\;\;\alpha},\zeta^{\alpha})^z$, with $z$ a label, as described in section~\ref{Hypermultiplet sec}. These family of hypers are not independent from each other as there are constraints $\Gamma=0$ that relate hypers $(A_i^{\;\;\alpha},\zeta^{\alpha})$ to $(A_i^{\;\;\alpha},\zeta^{\alpha})^z$, $(A_i^{\;\;\alpha},\zeta^{\alpha})^z$ to $(A_i^{\;\;\alpha},\zeta^{\alpha})^{(zz)}$ and so on. However, once these constraints are solved we are left with $(A_i^{\;\;\alpha},\zeta^{\alpha},A_i^{\;\;\alpha\,(z)})$ as the only independent fields. In order to localize the hypermultiplet fields we would look at the BPS equations
\begin{eqnarray}
&&\delta\zeta^{}=0\nonumber\\
&& \delta\zeta^{(z)}=0\nonumber\\
&& \delta\zeta^{(zz)}=0\nonumber\\
&&\ldots
\end{eqnarray}for a particular supercharge generated by $\delta$ under the constraints $\Gamma=0$. Naively this is a very complicated problem that we want to avoid since the constraint equations are complicated Klein-Gordon and Dirac type of equations. What we will do instead is to determine the solutions of the off-shell $\delta\zeta=0$ BPS equation for all eight Killing spinors and interpret the solution as an off-shell background in the hyper sector. In this case we are left with the BPS equation for the independent fields that can be easily solved to give
\beq
F_i^{\;\;\alpha}=-i\frac{2g}{\sqrt{8\pi G}}\sigma^{\alpha}_{3i}(H\cdot P),
\qquad
2g (J\cdot P)=-\frac{1}{L}
\eeq

\section{Evaluation of the  Action \label{app:Action}}

To compute the vector multiplet action (\ref{vectoraction}) we need the $N_{IJ}$ tensor, 
which is
\bal
N_{00}&=\frac{i}{8}\left(\frac{J^1}{J^0}\right)^\frac{3}{2}\left(t^3+\bar t^3\right),\quad %\\
N_{01}&=-\frac{3i}{8}\left(\frac{J^1}{J^0}\right)^\frac{1}{2}\left(t+\bar t\right),\quad
N_{11}=-\frac{3i}{8}\left(\frac{J^1}{J^0}\right)^{-\frac{1}{2}}\left(\frac{1}{t}+\frac{1}{\bar t}\right)
\eal
where we defined 
$t=\sqrt{\frac{X^1}{J^1}/\frac{X^0}{J^0}}=\sqrt{\frac{i+h^{1}/r}{i+h^0/r}}$
and  $\bar{t}$ its complex conjugate \footnote{We have chosen the branch cut of $\sqrt{X^1/X^0}$ along the negative real line and $\sqrt{-1}=-i$.This choice is justified by the fact that we are taking $J^0>0$ and $J^1<0$.}. Using $r=\cosh(\eta)$ we can write the off-shell action 
as
\bal
\label{vec-action}
S_\text{vec}&=\Omega_3L^2\frac{\sqrt{J^0(J)^3}}{2i}\int dr(r^2-1)\bigg[
-\left(1+\frac{(h^0)^2}{r^2}\right)\left(t+\bar t\right)^3
+\frac{3}{4}(h^{1})^2\left(\frac{1}{t}+\frac{1}{\bar t}\right)\frac{r^2-1}{r^4}
\\&
\quad+\frac{3}{2}h^{1}h^0\left(t+\bar t\right)\frac{r^2-1}{r^4}
-\frac{1}{4}(h^0)^2\left(t^3+\bar t^3\right)\frac{r^2-1}{r^4}
-\frac{3}{4}\left(\frac{1}{t}+\frac{1}{\bar t}\right)\left(1+i\frac{h^{1}}{r^2}\right)^2
\\&\quad-\frac{3}{2}\left(t+\bar t\right)\left(1+i\frac{h^{1}}{r^2}\right)\left(1+i\frac{h^0}{r^2}\right)
+\frac{1}{4}\left(t^3+\bar t^3\right)\left(1+i\frac{h^0}{r^2}\right)^2\bigg]
\eal
This turns out to be a total differential of a relatively simple function, so integrating over $r$ gives
\bal
\Omega_3L^2\frac{\sqrt{J^0(J)^3}}{2i}
\bigg[&\frac{(r-1)^2}{r}\sqrt{\frac{1+ih^{1}/r}{1+ih^0/r}}(-ih^{1}(1-2ih^0-r)-2r(2+r)-ih^0(3+r))+
\\&
+\frac{(r+1)^2}{r}\sqrt{\frac{1-ih^{1}/r}{1-ih^0/r}}(-ih^{1}(1-2ih^0+r)+2r(2-r)-ih^0(3-r))\bigg]
\eal

We now turn to the hypermultiplet action (\ref{hyperaction}).
We compute separately the quantities
\begin{eqnarray}
&&4g^2A^i_{\;\;\beta}\bar{X}^{\alpha}_{\;\;\gamma}X^{\gamma}_{\;\;\delta}A_i^{\;\;\delta}d_{\alpha}^{\;\;\beta}=\frac{g^2}{\pi G}(\bar{X}\cdot P)(X\cdot P)=\frac{g^2}{\pi G}\frac{(H^IP_I)^2}{\cosh^2(\eta)}+\frac{g^2}{\pi G}(J\cdot P)^2\\
&&gA^i_{\;\;\beta}Y^{jk\alpha}_{\;\;\;\gamma}A_k^{\;\;\gamma}\epsilon_{ij}d_{\alpha}^{\;\;\beta}=-\frac{gi}{2\pi G L}\frac{H^IP_I}{\cosh^2(\eta)}-\frac{g}{2\pi G L}(J\cdot P)\\
&&F_i^{\;\;\alpha}F^i_{\;\;\beta}d_{\alpha}^{\;\;\beta}=-\frac{g^2}{\pi G}\frac{(H^IP_I)^2}{\cosh^2(\eta)}.
\end{eqnarray}
Plugging these values in (\ref{hyperaction}) we obtain
\bal
\label{reg-hypers}
S_\text{hyp}&=-i\Omega_3 L^4\int^{r_0}_{1} dr(r^2-1)\frac{1}{r^2}\frac{g}{2\pi G L}(h^0J^0P_0+h^{1}J^{1}P_1)
\\&
=-i\frac{\Omega_3 g L^3}{2\pi G}(r_0-2)(h^0J^0P_0+h^{1}J^{1}P_1)+\mathcal{O}(1/r_0)
\eal
Using the attractor equations (\ref{attract values}) we get 
\beq
\displaystyle S_\text{hyp}=i\frac{\Omega_3L^2}{16\pi G}(r_0-2)(h^0+3h^{1})+\mathcal{O}(1/r_0)
\eeq

\section{Fayet-Illiopoulos terms on rigid $AdS_4$ background}

As we have seen that   the only role of the hypermultiplets is to provide a cosmological constant via their coupling to the vector multiplet. They play essentially no role in determining the localization locus and their contribution to the renormalized action is also minimal. In this section we discuss a somewhat different possible route to deriving the renormalized action without introducing the hypermultiplets or the flux boundary terms. 

Consider for this purpose a supersymmetric QFT on the $AdS_4$ background with Fayet-Illioupoulos terms. The purpose of this exercise is to show that one can obtain the renormalized action is given by  (\ref{Ren S}).  
The setup is precisely the same as before but with additional FI terms. In order to ensure supersymmetry, the scalars must be conformally coupled to the curvature of the background and this is what superconformal gravity precisely does. Since the theory is still off-shell the localization solutions are still the same except that the boundary conditions for the scalars must be given.

We therefore consider a theory of vectors conformally coupled to $AdS_4$
\beq\label{vecs on ads4}
S=\int N_{IJ}\bar{X}^IX^J\frac{R}{6}+N_{IJ}\partial\bar{X}^I\partial X^J-\frac{1}{8}N_{IJ}Y^{ij I}Y^J_{ij},
\eeq where we just show the bosonic part, and the curvature $R$ is now fixed to $-12/L^2$. As noted before the coupling of the vectors to the hypermultiplets looks very much like a FI coupling since the auxiliary fields $Y^i_{\;\;j}$ couple linearly. On $AdS_4$ we can show that the combination
\beq
Y^{i}_{\;\;j}\sigma^{j}_{3\;\; i}-\frac{4}{L}(X-\bar{X})
\eeq varies by supersymmetry into a total derivative and therefore it can be added to the action (\ref{vecs on ads4}). We tune the FI couplings $\zeta_I$ in order to obtain the same equations of motion as in (\ref{EOM}), that is,
\beq
\zeta_I=-\frac{gi}{8\pi G}P_I.
\eeq

The localization equations are not changed and therefore we find the same instanton solutions. We compute the renormalized action to get
precisely the same renormalized action as (\ref{ren-action}). In the gravitational context there was an additional   flux boundary term but with the FI couplings it is not required.

\bibliographystyle{JHEP}
\bibliography{ads4}

\end{document}